\documentclass{IEEEtran}
\usepackage{amsmath,amssymb,amsfonts}
\usepackage{algorithmic}
\usepackage{graphicx}
\usepackage{textcomp}

\usepackage[table,xcdraw]{xcolor}
\usepackage{pdfpages}
\usepackage{svg}
\usepackage{mathtools}
\usepackage{caption}
\usepackage{subcaption}
\usepackage{blkarray, bigstrut}
\usepackage[backend=biber,
            style=numeric,
            abbreviate=false,
            dateabbrev=false,
            alldates=long,
            maxnames=10
            ]{biblatex}

\usepackage{supertabular,booktabs}
\usepackage{multicol}
\usepackage{float}
\usepackage{multirow}
\usepackage{titlesec}

\usepackage{diagbox}

\usepackage[per-mode = symbol]{siunitx}   

\def\BibTeX{{\rm B\kern-.05em{\sc i\kern-.025em b}\kern-.08em
    T\kern-.1667em\lower.7ex\hbox{E}\kern-.125emX}}

\definecolor{BluePP}{cmyk}{0.36, 0.34, 0, 0.04}
\definecolor{YellowPP}{cmyk}{0,0.07, 0.33, 0.02}
\definecolor{RedPP}{cmyk}{0, 0.30, 0.3, 0.2}
\definecolor{PurplePP}{cmyk}{0,0.17, 0.06, 0.2}
\definecolor{OrangePP}{cmyk}{0, 0.26, 0.39, 0.01}
\definecolor{GreenPP}{cmyk}{0.35, 0, 0.15, 0.09}

\begin{document}

\title{Quantitative modeling and simulation of biochemical processes in the human body}

\author{Jacob Bendsen, Peter Emil Carstensen, Asbjørn Thode Reenberg, Tobias K. S. Ritschel, \\ John Bagterp Jørgensen
\thanks{J. Bendsen, P.E. Carstensen, T.K.S. Ritschel and J.B. J{\o}rgensen are with Department of Applied Mathematics and Computer Science, Technical University of Denmark, Kgs Lyngby, Denmark. Corresponding author: J.B. J{\o}rgensen (e-mail: jbjo@dtu.dk). J. Bendsen and P.E. Carstensen contributed equally to the paper.}
}

\maketitle

\begin{abstract}
We present a whole-body model of human metabolism that utilizes a system of organs and blood vessels to simulate the enzymatic reactions. The model focuses on key organs, including the brain, heart and lungs, liver, gut, and kidney, as well as muscle and adipose tissue. The model equations are formulated using stoichiometry and Michaelis-Menten kinetics to describe the enzymatic reactions. We demonstrate how the model can be used to simulate the effects of prolonged fasting and intermittent fasting on selected metabolite concentrations and glucose flux. Furthermore, by simulating intermittent fasting the effect on the carbohydrate, the protein and the lipid storage is examined. We propose this method as a simple and intuitive approach for modeling the human metabolism, which is general, systematic and easy to incorporate. This could have potential applications in PK/PD drug development and in understanding metabolic disorders.
\end{abstract}


\section{Introduction}
\label{sec:introduction}

The human body is comprised of various metabolic processes that continuously form and break down metabolites. The body's metabolism operates in a systematic manner to keep the organism alive, and this concept is applied in whole-body modeling, which views the body as a unified unit \cite{miesfeld_mcevoy_2017}. This approach allows the prediction of metabolite concentrations in specific organs, which is relevant e.g. to PK/PD drug development \cite{derendorf_schmidt_rowland_tozer_2020}.

There are different methods for modeling the human metabolism, depending on the intended use of the model. In our approach, we consider the system of organs and blood vessels as a whole-body model, with enzymatic reactions taking place in the organs. Flux between the organs is included by connecting them through the blood vessels. Different reactions and reaction rates are defined based on the role of each organ and their metabolism. Following this approach, it is possible to simulate the metabolism of man under various conditions.

Current literature includes whole-body models with varying levels of complexity. For example, \cite{sorensen_1978} improved earlier, inadequate models by focusing on glucose, insulin, and glucagon dynamics using a simple whole-body model. More complex models have since been developed, such as \cite{panunzi_pompa_borri_piemonte_gaetano_2020}, which extends \cite{sorensen_1978} by incorporating food intake. Other authors have expanded on whole-body models, incorporating multiple metabolites through stoichiometry \cite{kim_saidel_cabrera_2006,dash_li_kim_saidel_cabrera_2008, KURATA2021102101}. However, a simple and intuitive mathematical approach for formulating these models is not readily available. 

In this paper, we expand on our previous work \cite{Carstensen_Bendsen_et_al} in which we  describe a whole-body model of the human metabolism and present a general, systematic, and intuitive method for formulating model equations. The five key organs considered in the model are the brain, heart and lungs, liver, gut, and kidney, with muscle tissue and adipose tissue each simplified as a single compartment. The metabolic processes within the organs are explained through the stoichiometry of enzymatic reactions, which are described using Michaelis-Menten kinetics. We simulate the feed-fast cycle to study the impact of prolonged fasting on selected metabolite concentrations and glucose flux, regular food intake to examine longer cyclic behavior, and intermittent fasting to examine its effect on carbohydrate, protein, and lipid storage.

The remaining part of this paper is structured as follows. Section II describes the approach for whole-body modeling. Section III considers the biology of macronutrient metabolism, which is formulated as a model in section IV using the mathematical approach. Section V presents the simulation results. Our formulated model and assumptions are discussed in section VI. Finally, section VII concludes on our findings.

\section{Mathematical approach}
\label{sec:mathmaticalmodel}

The general model is described as a system in which metabolites flow in, are metabolized, and flow out. The dynamics of a single compartment is defined by the general differential equation
\begin{equation}
\label{eq:general_eq}
    V \frac{dC}{dt} = M (Q_{in}C_{in} - Q_{out}C) + RV , 
\end{equation}
where $V$ is the volume, $C$ is a vector containing the concentration of the metabolites, $M$ is the external and internal component ordering, $Q_{in}$ is the flow rate of what goes in, $C_{in}$ is a vector containing the concentration of the metabolites that flow in, $Q_{out}$ is the flow rate of what goes out and $R$ is the production rates. The compartments are coupled through concentration gradients in the blood vessels that connect the compartments. $M$ is a square matrix containing only ones and zeros in the diagonal corresponding to the metabolites distributed through the blood vessels (circulating metabolites). For instance, the circulating metabolite, $C_i$, corresponds to $M_{i,i}=1$. The production rate $R$ is incorporated as a vector defined by
\begin{equation}
    R = (T S)' T r ,
\end{equation}
where $T$ is a matrix of reactions that occur, $S$ is a stoichiometric matrix containing all reactions and $r$ is a vector with the kinetics for the reactions. $T$ contains ones and zeros corresponding to which reactions from the stoichiometric matrix, that are present in the compartment. For instance, a compartment which involves reaction $1,3$ and $5$ from the stoichiometric matrix have $T_{1,1} = 1, T_{2,3} = 1, T_{3,5} = 1$, and zeros elsewhere. The reaction rate vector, $r$, is a function of the concentration of each metabolite:
\begin{equation}
    r = r(C).
\end{equation}
To utilize \eqref{eq:general_eq}, in a whole-body model, it must be formulated for each compartment.

\subsection{Example model}

We present an example of the model's method and logic, in a simple three compartment model ($k$) with six metabolites ($i$) and six reactions ($j$), resulting in 3 differential equations and stoichiometric matrix 6$\times$6. We define this the example model. This results in the following sets:
\begin{align*}
k \in K & =  \{H,L,G \} \\
i \in I & =  \{A, B, C, D, E, F \} \\
j \in J & =  \{r_1, r_2, r_3, r_4, r_5, r_6 \}
\end{align*}

\begin{figure}[tb]
    \centering
    \includegraphics[width=0.87\columnwidth]{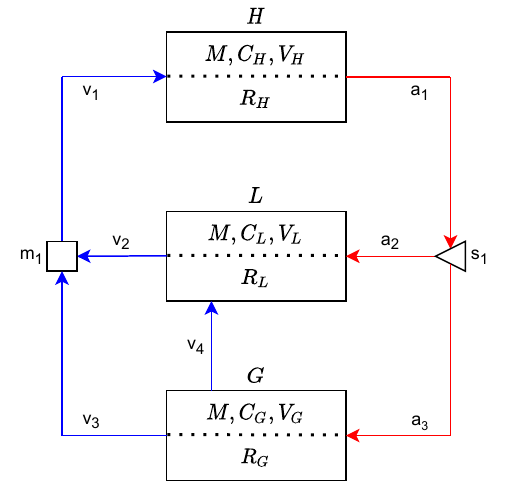}
    \caption{Schematic representation of the example model. Solid arrows represent the blood circulation. $M,C_k,V_k$ represents the blood tissue exchange and $R_k$ represents the reactions happening inside the cell. The dotted lines in the compartments suggests free diffusion, as cell-permeability is not included.}
    \label{fig:example_flow}
\end{figure}

The blood circulation can be split into two parts, the arteries (red) and the veins (blue). The total blood flow, $Q$, is at all times preserved, and the local blood flow can be calculated using mixers and splitters. They are either explicitly modelled as organs, or implicitly as shown by the triangle (splitter) or square (mixer). A splitter divides the blood flow: $a_1 = a_2 + a_3$, and a mixer combines: $v_2+v_3 = v_1$. The same is true for organs. From the flow diagram, the \textit{L}-compartment is a mixer such that: $v_4+a_2 = v_2$ and the \textit{G}-compartment is a splitter such that: $a_3 = v_4+v_3$. As blood flow is preserved, the flow coming into the \textit{H}-compartment is the same as the flow coming out of the \textit{H}-compartment. \\

The differential equations describing the mass conservation for each organ are all in the form \eqref{eq:general_eq}.

\begin{subequations}
\begin{align}
    V_H\frac{dC_H}{dt} & = M (Q_{LH} C_L + Q_{GH} C_G - Q_H C_H) + R_H V_H \\
    V_L\frac{dC_L}{dt} & = M (Q_{GL} C_G + Q_L C_H - Q_{LH} C_L) + R_L V_L\\
    V_G\frac{dC_G}{dt} & = M (Q_G C_H - Q_{GL} C_G - Q_{GH} C_G) + R_G V_G
\end{align}
\end{subequations} \\
The subscript $Q_{LH}$ is the blood flow from the liver to the heart, and can be derived as $Q_{LH} = v_2 = a_2+v_4$. \\

\begin{figure}[tb]
    \centering
    \includegraphics[width=0.8\columnwidth]{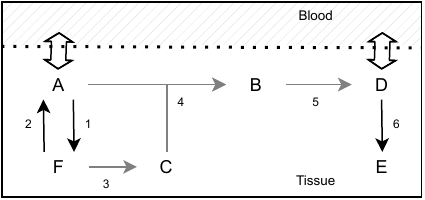}
    \caption{Diagram of the metabolic pathways in a cell. The hollow double-sided arrows reflect that the metabolite is distributed through the blood. The black arrows indicate reactions that happen in all compartments. The grey arrows indicate reactions that happen in some compartments. Reactions are numbered corresponding to the stoichiometric matrix in equation (\ref{eq:stoi_example}).}
    \label{fig:example_metabolic_map}
\end{figure}

As this model utilizes a stoichiometric matrix to model cell metabolism \cite{yasemi_jolicoeur_2021}, we include the addition of $M$ and $R_k$. $M$ is a square matrix containing only ones and zeros in the diagonal corresponding to the circulating metabolites. In this example $A$ and $D$ are the circulating metabolites. $R_k$ is the production rates of metabolites inside the tissues. The reactions occurring inside each of the compartments in Fig. \ref{fig:example_flow} are shown in Fig. \ref{fig:example_metabolic_map} and Table \ref{tab:example_stoichio_table} \\

\setlength\extrarowheight{-4pt}
\begin{table}[tb]
\centering
 \caption{Summary of the stoichiometric reactions and its kinetics in the example model.}
 \label{tab:example_stoichio_table}
 \resizebox{0.75\columnwidth}{!}{
  \begin{tabular}{*{3}{c}}
    \# R  & Stoichiometric   & Kinetic \\[0.5ex]
   \midrule
    1 &  $A \rightarrow F$  &  $r_1 = p_1 C_A$  \\
   \midrule
    2 &  $F \rightarrow A$  &  $r_2 = p_2 C_F$ \\
   \midrule
    3 &  $F \rightarrow C$  &  $r_3 = p_3 C_F $  \\
   \midrule
    4 &  $A + C \rightarrow B$  &  $r_4 = p_4 C_A C_C$ \\
    \midrule
    5 &  $B \rightarrow 2 D$  &  $r_5 = p_5 C_B$  \\
    \midrule
    6 &  $D \rightarrow E$  &  $r_6 = p_6 C_D$  \\
   \bottomrule
  \end{tabular}
  }
\end{table}

\setlength\extrarowheight{0pt}

$r$ is a reaction rate, that depends on the concentration $C$ and a scalar $p$.

To calculate $R_k$ in the three equations, the stoichiometric matrix is defined based on Table \ref{tab:example_stoichio_table}  \\

\begin{equation}
\label{eq:stoi_example}
S =
\begin{blockarray}{l c c c c c c}
\begin{block}{l c c c c c c}
& A & B & C & D & E & F & \\
\end{block}
\begin{block}{l [c c c c c c]}
r_1 & -1 & 0  & 0  & 0  & 0 & 1 \\
r_2 & 1  & 0  & 0  & 0  & 0 & -1 \\
r_3 & 0  & 0  & 1  & 0  & 0 & -1 \\
r_4 & -1 & 1  & -1 & 0  & 0 &  0 \\
r_5 & 0  & -1 & 0  & 2  & 0 & 0 \\
r_6 & 0  & 0  & 0  & -1 & 1 & 0 \\
\end{block}
\end{blockarray},
\end{equation}

where rows are the coefficients of the reactions and columns are metabolites. To find the changes in concentrations in each compartment we must define a compartment specific matrix, that defines which reactions happens in each cell. To do this we use the matrix of the occuring reactions, $T_k$. In this example we define $T_H$ in the \textit{H}-compartment where only reactions 1, 2 and 6 occur,

\begin{equation}
    T_H = 
    \begin{bmatrix}
        1 & 0 & 0 & 0 & 0 & 0 \\
        0 & 1 & 0 & 0 & 0 & 0 \\
        0 & 0 & 0 & 0 & 0 & 1
    \end{bmatrix},
\end{equation}

we can then find the changes in the production rate of the \textit{H}-compartment vector $R_H$ as

\begin{equation}
    R_H =  (T_HS)'T_Hr_H .
\end{equation}

Where reaction kinetics can be described dependant on the reactant being used and an equation describing enzyme kinetics. These equations could e.g. be Hill equations or Michaelis-Menten kinetics. Further, several additions could be made to include reaction specific equations, such as hormonal effect or capacity-limited reactions. \\

\section{Biological considerations}

Given the definition of the mathematical framework, we now define the key biological considerations, that we make use of to showcase our approach. \\

The organs of the human body are connected through our blood vessels. They are thus able to transport nutrients to one another. The blood flows from the heart to all other organs in the arteries, and flows through the veins back to the heart, completing the blood circulation. This process will e.g. allow the liver to produce glucose for the brain to use as a fuel. \\

\begin{figure}[tb]
    \centering
    \includegraphics[width=0.85\columnwidth]{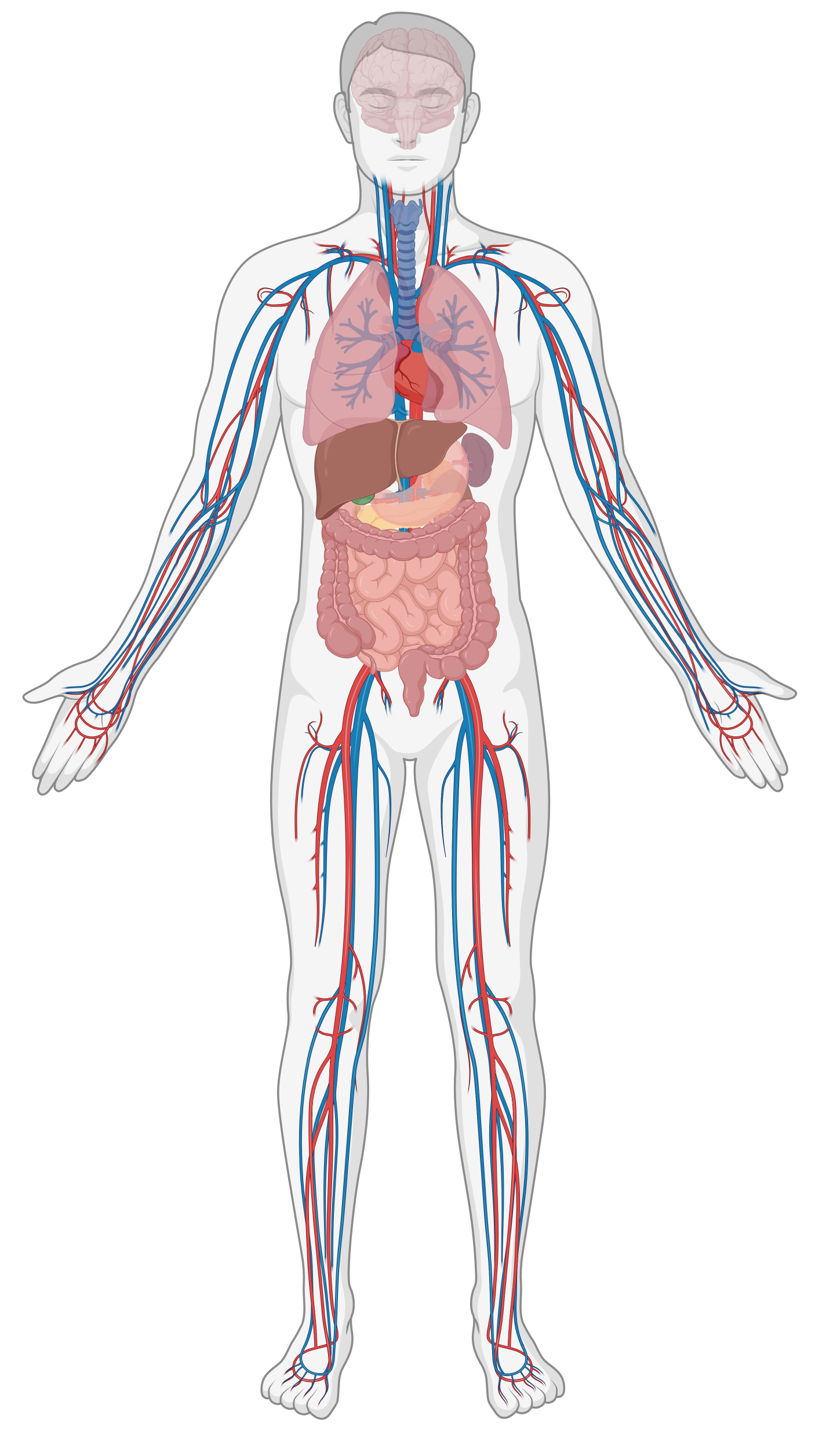}
    \caption{Qualitative whole-body model of the blood circulation. Arteries: red. Veins: blue. Created with BioRender.com.}
    \label{fig:whole-body}
\end{figure}

\begin{figure}[tb]
    \centering
    \includegraphics[width=\columnwidth]{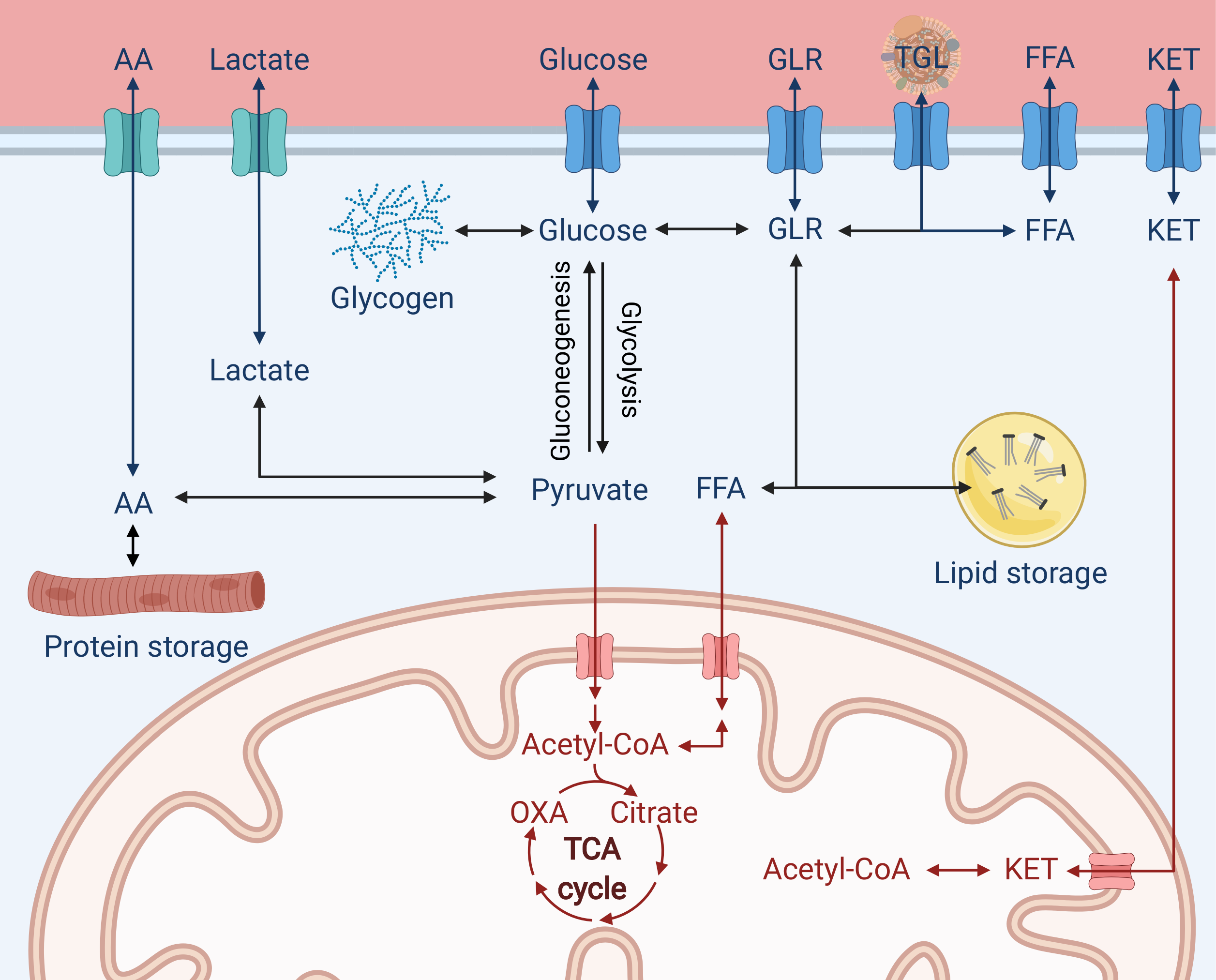}
    \caption{Qualitative model of a metabolic system, with the following circulating metabolites: amino acids (AA), lactate (LAC), glucose (GLC), glycerol (GLR), triglycerides (TGL), free fatty acids (FFA) and ketone bodies (KET) in the top part of the figure. Multiple reactions can occur in a single arrow, e.g. glycolysis. Created with BioRender.com.}
    \label{fig:Qualitative_cell}
\end{figure}

Many different types of cells are present in the organs. In each of these cells, a complex metabolic network allows for the metabolism of nutrients. In order to gain an understanding of the metabolic network, a simplified qualitative model is presented in Fig. \ref{fig:Qualitative_cell}, showing the overall metabolic processes in a given eukaryotic cell. \\

Further, as we include lipid and protein storage, we consider the 4 different stages from fed to starvation: 1) postpandrial, 2) postabsorptive, 3) fasting, and 4) starvation \cite{gropper_smith_carr_2018}. The metabolite concentrations qualitatively drastically differ in each of these stages. 

The postprandial phase is generally characterized by a high insulin-to-glucagon ratio, which stimulates the uptake and storage pathways found in liver, muscle and adipose tissue. The postabsorptive state is catabolic, i.e. glycogenolysis and gluconeogenesis is favoured in the liver to maintain blood glucose levels. In the postabsorptive stage, we expect to see a constant glucose concentration, with declining glycogen concentrations \cite{gropper_smith_carr_2018}, due to the postabsorptive metabolism being characterized by a higher glucagon-to-insulin ratio. 

In the fasting stage, the blood glucose concentration begins to fall slowly, as glycogen depots are further depleted, and gluconeogenesis cannot maintain constant glucose levels. The glucagon-to-insulin ratio further increases, facilitating gluconeogenesis. Additionally, amino acid levels increase as the breakdown of muscle proteins is further facilitated for use in gluconeogenesis. The lipolysis in adipose tissue intensifies to provide fatty acids for fuel and glycerol for gluconeogenesis. 

When fasting increases beyond 48 hours, it is characterized as starvation. A major shift in metabolic fuels occurs, as the body tries to preserve proteins. Lipolysis rates accelerates, and fatty acids are the main source of energy in most tissues. As the brain cannot utilize fatty acids, the ketone body formation is substantially increased in the liver. The ketone bodies becomes the main fuel for the brain, as well as being utilized by heart and skeletal muscle for energy.

\section{Model}
\label{sec:model}

\begin{figure}[tb]
    \centering
    \includegraphics[width=0.82\columnwidth, keepaspectratio]{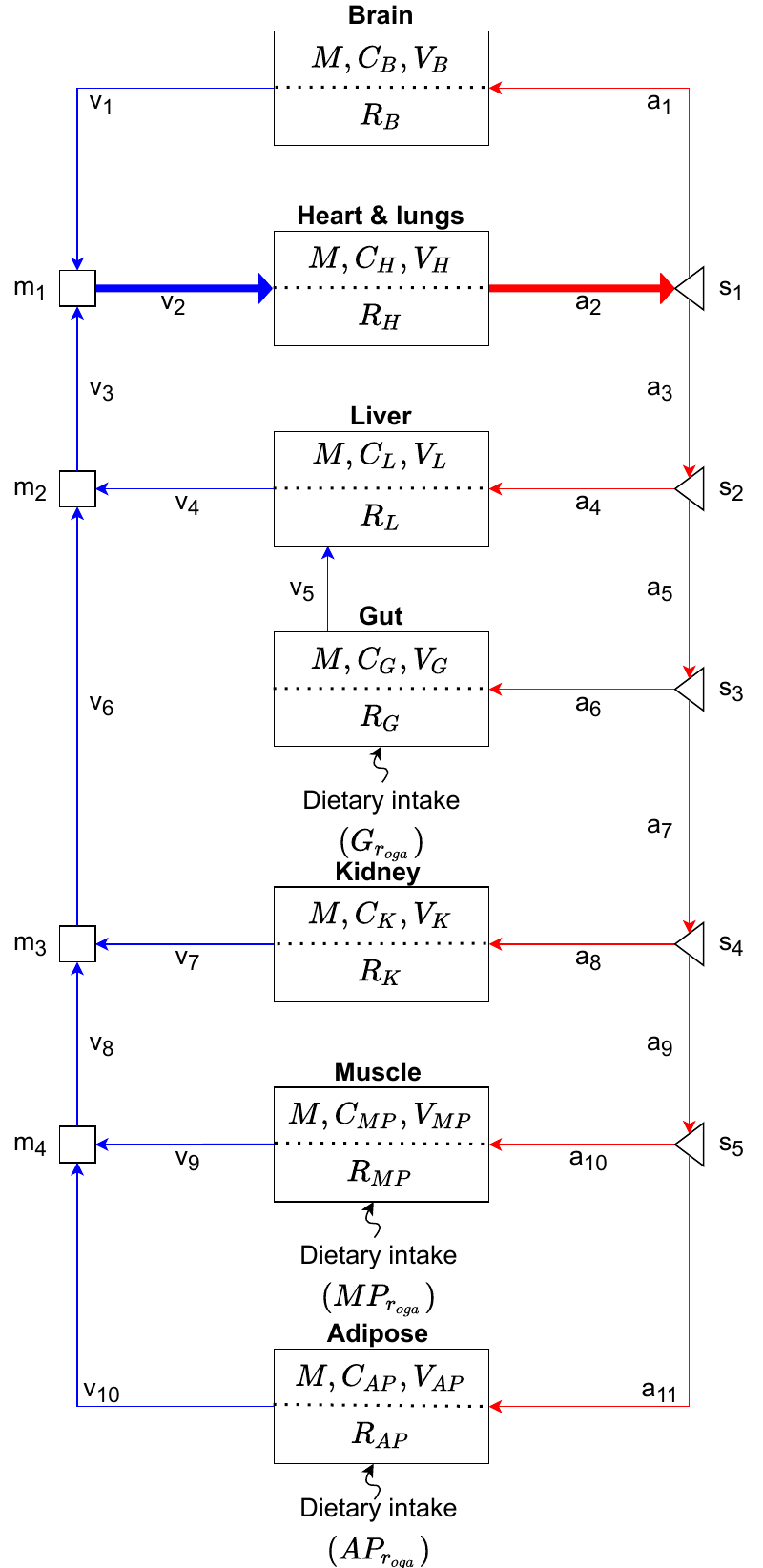}
    \caption{Schematic representation of the whole-body model. Solid arrows represent blood circulation, where the right side is the arteries and the left side the veins.
    Thick arrows, $a_2$ and $v_2$, represents joining of flows from other organs.
    $M$,$C_k$,$V_k$ represents the blood tissue exchange and $R_k$ represents the reactions happening inside the cell. The dotted lines in the compartments suggest free diffusion, as cell-permeability is not included.}
    \label{fig:Bends_cars_flow_diagram}
\end{figure}

We now present a model containing 7 compartments and 16 metabolites split into 31 reactions including the hormonal effect from two signal molecules, insulin and glucagon, on specific tissues. Fig. \ref{fig:Bends_cars_flow_diagram} shows a flow diagram of the whole-body model. 

\begin{figure}[tb]
    \centering
    \includegraphics[width=\columnwidth]{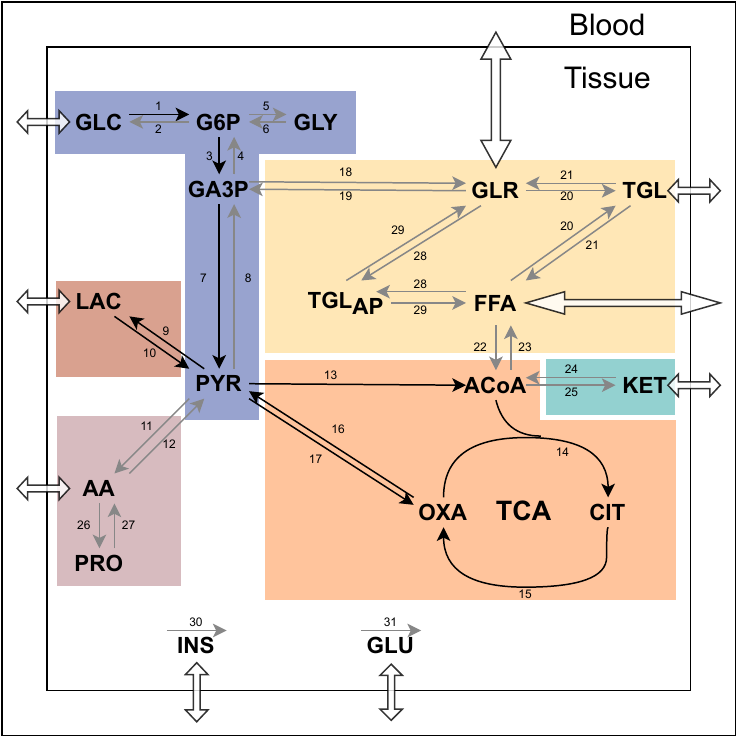}
    \caption{Diagram of metabolic pathways in cell. The hollow double-sided arrows indicate that the metabolite is distributed through the blood. The black arrows indicate reactions that happen in all organs. The grey arrows indicate reactions that only happen in some organs. Metabolites are shown as 2-4 letter abbreviations. The stoichiometric matrix is formed from the numbering of the reactions.}
    \label{fig:metabolic_map}
\end{figure}

The model is created using the methodology outlined in Section \ref{sec:mathmaticalmodel}, to depict the energy metabolism of proteins, carbohydrates, and lipids. The major biochemical pathways involved in these macronutrients were included to simulate their behavior under different conditions. The addition of insulin and glucagon maintain stability and prevent large transient periods in the simulation post-food intake, as these hormones play a crucial role in anabolism and catabolism. As seen from Fig. \ref{fig:metabolic_map}, many metabolic pathways are specific to certain organs. Out of 31 reactions, 10 occur in all organs, while the remaining 21 are tissue-specific due to the specialized role of each organ. An overview of metabolic pathways in all organs is provided in Table \ref{tab:summary_pathways}.

Insulin and glucagon secretion/clearance is incorporated as a single reaction for simplicity, as they are not transformed into or derived from any included metabolite. Their production rates were adopted from \cite{sorensen_1978}, as described in Section \ref{sec:hormonal_model}.

Due to the variation between individuals, the most accurate mathematical equations to describe each of the 31 metabolic reactions may not be known. In the model, we use Michaelis-Menten kinetics to describe the reaction rates. This approach is particularly useful in describing enzymatic reactions, as the concentrations of enzymes impose an upper limit on their reaction rate. For reactions with two substrates, we utilize the two substrate Michaelis-Menten kinetics from \cite{MM_2_reactants}. Alternatives to Michaelis-Menten kinetics include adjusting the mathematical formula in the reaction rate vector, such as positive hyperbolic tangent functions in \cite{sorensen_1978} or simple first-order kinetics in \cite{kim_saidel_cabrera_2006}. Exclusion of a reaction from an organ does not mean it never occurs in reality, but rather that it is excluded for simplicity. An example is glycogen formation, which also occurs in the brain, heart, and adipose tissue, but in such small amounts that it becomes negligible \cite{gropper_smith_carr_2018}.

\begin{table}[tb]
    \centering
    \caption{Summary of the metabolic pathways in all tissues. Organs are in the columns and reactions are in the rows. Grey squares represent, that the reaction is present in the organ, white squares indicate the reaction is disregarded. The reactions can be found according to their number in appendix \ref{tab:appendix_reaction}}
    \label{tab:summary_pathways}
    \includegraphics[width=\columnwidth]{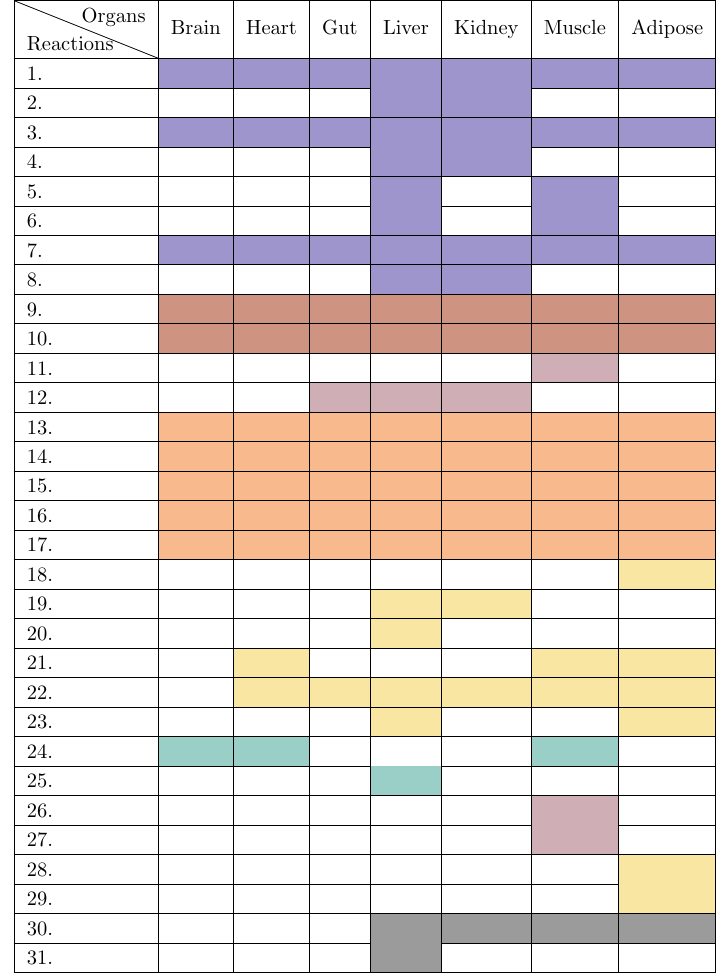}
\end{table}

\subsection{Inclusion of a hormonal model}
\label{sec:hormonal_model}
Sorensen \cite{sorensen_1978} created a straightforward glucagon model that captures pancreatic glucagon release, and found that a one compartment model was sufficient to describe glucagon release and clearance. Additionally, Sorensen \cite{sorensen_1978} designed an insulin model described by a six compartment model. Instead of including the glucagon model as a single compartment model and insulin as a six compartment model, like \cite{panunzi_pompa_borri_piemonte_gaetano_2020} and \cite{sorensen_1978}, we use a seven compartment model that encompasses insulin and glucagon. This allows us to calculate their production rates by including insulin and glucagon as metabolites in the stoichiometric matrix.

\setlength\extrarowheight{3pt}
\begin{table}[tb]
\centering
\caption{Reactions affected by insulin and glucagon \cite{gropper_smith_carr_2018, miesfeld_mcevoy_2017}. $\Uparrow$ symbolizes increased stimulation and $\Downarrow$ symbolizes decreased stimulation.}
\label{tab:insulin_glucagon_reactions}
\resizebox{\columnwidth}{!}{%

\begin{tabular}{lcccc}
    \textbf{Hormone}  & \textbf{Effect} & \textbf{\# R} & \textbf{Reaction} & \textbf{Affected Organs} \\[0.5ex] \hline
    \textbf{Insulin}  & $\Uparrow$ & 1 & GLC $\rightarrow$ G6P  & Liver, muscle, adipose tissue \\ \cmidrule{3-5}
                    &  $\Uparrow$ & 3 & G6P $\rightarrow$ 2 GA3P  & Liver, muscle tissue \\ \cmidrule{3-5} 
                    &  $\Uparrow$ & 5 & G6P $\rightarrow$ GLY  & Liver, muscle tissue \\ \cmidrule{3-5} 
                    & $\Uparrow$ & 13 &  PYR $\rightarrow$ ACoA  & Liver, muscle tissue \\ \cmidrule{3-5} 
                    & $\Uparrow$ & 21 & TGL $\rightarrow$ 3 FFA + GLR  & Adipose tissue \\ \cmidrule{3-5} 
                    & $\Uparrow$ & 23 & 7 ACoA $\rightarrow$ FFA  & Liver \\ \cmidrule{3-5} 
                    & $\Uparrow$ & 26 & AA $\rightarrow$ PRO   & Muscle tissue \\ \cmidrule{3-5} 
                    & $\Uparrow$ & 28 & 3 FFA + GLR $\rightarrow$ TGL$_\mathrm{AP}$   & Adipose tissue \\ \cmidrule{2-5} 
                    & $\Downarrow$ & 4 & 2 GA3P $\rightarrow$ G6P  & Liver \\ \cmidrule{3-5} 
                    & $\Downarrow$ & 6 & GLY $\rightarrow$ G6P    & Liver, muscle tissue \\ \cmidrule{3-5}
                    & $\Downarrow$ & 27 & PRO $\rightarrow$ AA    & Muscle tissue           \\ \cmidrule{3-5}
                    & $\Downarrow$ & 29 & TGL$_\mathrm{AP}$ $\rightarrow$ 3 FFA + GLR     & Adipose tissue \\ 
    \midrule 
    \textbf{Glucagon} & $\Uparrow$ & 4 &  2 GA3P $\rightarrow$ G6P & Liver \\ \cmidrule{3-5} 
                    & $\Uparrow$ & 6 & GLY $\rightarrow$ G6P & Liver \\ \cmidrule{3-5} 
                    & $\Uparrow$ & 29 &  TGL$_\mathrm{AP}$ $\rightarrow$ 3 FFA + GLR & Adipose tissue \\ \cmidrule{2-5} 
                    & $\Downarrow$ & 3 & G6P $\rightarrow$ 2 GA3P & Liver \\ \cmidrule{3-5}
                    & $\Downarrow$ & 5 & G6P $\rightarrow$ GLY & Liver \\  
   \bottomrule
  \end{tabular}%
}
\end{table}
\setlength\extrarowheight{0pt}

Table \ref{tab:insulin_glucagon_reactions} illustrates the reactions that are influenced by insulin and glucagon, along with the affected organ. Although the qualitative impact of insulin and glucagon is known \cite{miesfeld_mcevoy_2017}, the specific model parameters for insulin and glucagon are not always known. Based on the reciprocal relationship between insulin ($I$) and glucagon ($\Gamma$), we use a simple function that considers their ratio to determine their impact on the system and adjust the reaction rates in Table \ref{tab:insulin_glucagon_reactions} accordingly. If insulin has an impact on a reaction, the functions are
\begin{equation}
    \textbf{Insulin activation:} \left(\frac{I_k}{I_k^B}\right)^{\mu_j} V_{max_j} ,
\end{equation}
\begin{equation}
    \textbf{Insulin inhibition:} \left(\frac{I_k^B}{I_k}\right)^{\mu_j} V_{max_j} ,
\end{equation}
but if both insulin and glucagon have opposing effects on a reaction, the insulin-to-glucagon or glucagon-to-insulin ratio is used instead
\begin{equation}
    \textbf{Insulin-to-Glucagon stimulation:} \left(\frac{\Gamma_k^B}{\Gamma_k} \hspace{1mm} \frac{I_k}{I_k^B}\right)^{\mu_j} V_{max_j} ,
\end{equation}
\begin{equation}
    \textbf{Glucagon-to-insulin stimulation:} \left(\frac{\Gamma_k}{\Gamma_k^B} \hspace{1mm} \frac{I_k^B}{I_k}\right)^{\mu_j} V_{max_j} ,
\end{equation}
where $j$ is the reaction, e.g. \textit{GLC$\rightarrow$G6P}, $k$ is the compartment and the superscript $B$ indicates that it is the basal-value. $V_{max_j}$ is the maximum rate in the Michaelis-Menten kinetics and $\mu_j$ is a scaling parameter for the hormonal effect. These simple functions are equal to $V_{max_j}$ at steady state, which occurs when the blood glucose concentration is at $5\,\mathrm{mmol}/\mathrm{L}$, such that it is independent of $\mu_j$ in steady-state. This is an important property, since many of the parameters used in the reactions are estimated based on metabolite homeostasis. 

\subsection{Inclusion of a modified SIMO-model}

Our study adopts a modified version of the SIMO model \cite{panunzi_pompa_borri_piemonte_gaetano_2020} to account for the uptake of glucose, amino acids, and lipids from dietary intake. However, this model assumes a uniform rate for all macronutrients, which is not a realistic physiological representation as different macronutrients have distinct absorption rates \cite{miesfeld_mcevoy_2017}. Since data or mathematical models are not available to describe the specific uptake of macronutrients, we chose the SIMO model as the simplest option.

\begin{figure}[tb]
    \centering
    \includegraphics[width=0.9\columnwidth]{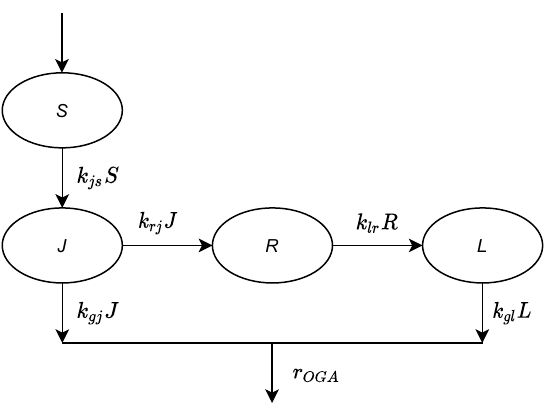}
    \caption{Schematic representation of the digestive tract. \textit{S} represents the amount of macronutrients in the stomach, \textit{J} the jejunum, \textit{R} a delay and \textit{L} the amount in the ileum. $r_{OGA}$ is a vector describing the uptake of the three macronutrients.}
    \label{fig:simo}
\end{figure}

The resulting uptake of macronutrients is represented by the equation:
\begin{equation}
    r_{oga} = k_{gj}J + k_{gl}L = \begin{bmatrix}
           k_{gj}J_{GLC} + k_{gl}L_{GLC} \\
           k_{gj}J_{AA} + k_{gl}L_{AA} \\
           k_{gj}J_{TGL} + k_{gl}L_{TGL}
         \end{bmatrix},
\end{equation}

where $r_{oga}$ is a 3$\times$1 vector, where the first two macronutrients, i.e. glucose and amino acids, are taken up by the gut. Triglycerides, however, enter the lymphatic system as chylomicrons and are transported to muscle and adipose tissue before reaching the blood circulation. The last instance of $r_{oga}$ is then delivered to muscle and adipose tissue, where a 50/50\% distribution is assumed in the two tissues. Since the SIMO model does not provide the specific uptake rate of amino acids and fat, we use the uptake rate $k_{gj}$ and $k_{gl}$ from glucose.

It is included in the model as the parameters:

\begin{align*}
    \text{Gut: } &  G_{r_{oga}} \\
    \text{Muscle: } & MP_{r_{oga}} \\
    \text{Adipose: } & AP_{r_{oga}}
\end{align*}

The modified SIMO model is incorporated as an additional input in the differential equations rather than directly into the production rate vector $R$. This is considered an extension to the general methodology, as the modified SIMO model provides inputs from meal consumption.

\section{Simulation results}
\label{sec:simulationresults}

\begin{figure}[tb]
    \centering
    \includegraphics[width=\columnwidth]{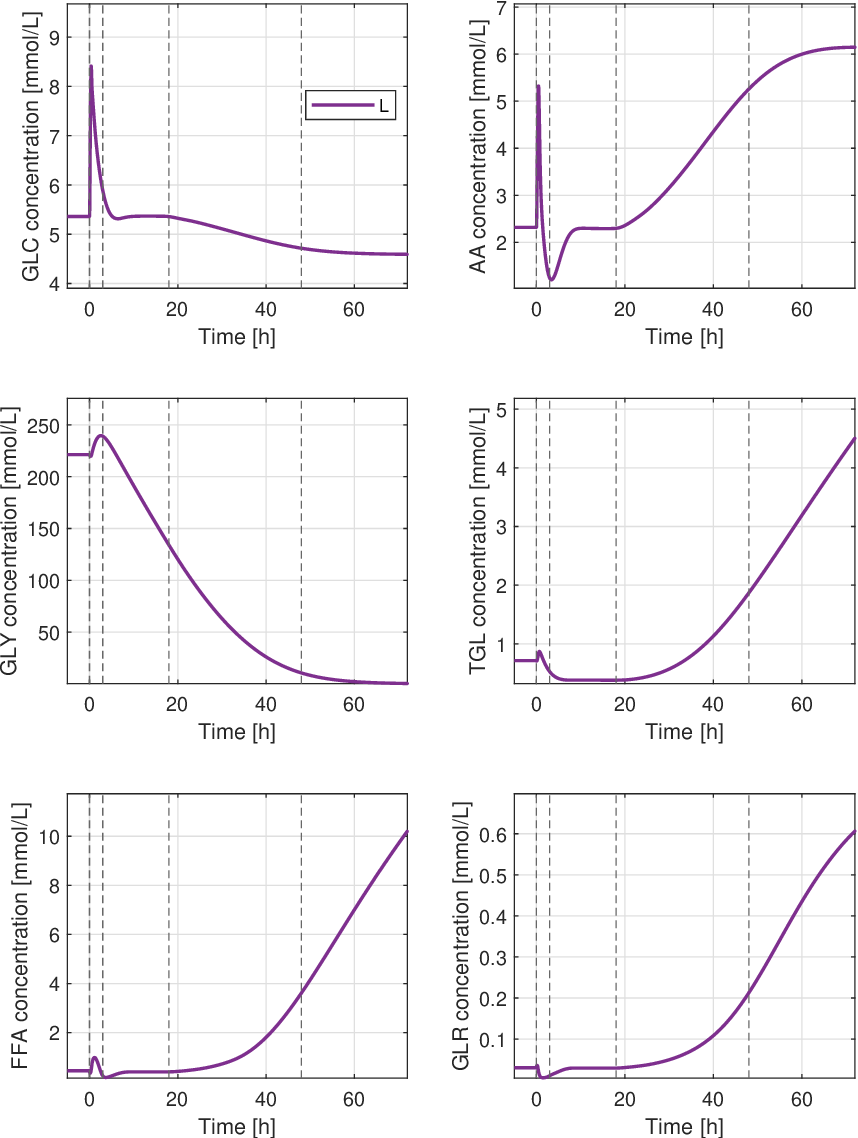}
    \caption{The metabolite concentrations of glucose ($GLC$), amino acids ($AA$), glycogen storage ($GLY$), triglycerides ($TGL$), free fatty acids ($FFA$) and glycerol ($GLR$) in the liver after an initial meal of 60 g glucose, 24 g protein and 16 g fat and an accompanying fasting for 72 hours. Vertical dotted lines indicate the four stages in the feed-fast cycle.}
    \label{fig:Liver_metabolites}
\end{figure}

We now simulate the model presented in Section \ref{sec:model} in MATLAB. We start with a single meal at steady-state and observe the changes in metabolite levels over the next 72 hours of simulated fasting and inactivity. Fig. \ref{fig:Liver_metabolites} displays the concentrations of glucose, amino acids, triglycerides, glycerol, free fatty acids, and glycogen. The graphs of $GLC$, $AA$, and $TGL$ illustrate the metabolites ingested through the modified SIMO model. After the meal, the glucose concentration initially rises and returns to baseline, staying constant for approximately 10 hours. It then decreases as glycogen storage diminishes. At the same time, other metabolites increase, especially glycerol, free fatty acids, and triglycerides as the simulated patient enters starvation and lipids become the main energy source for various organs. A rise in triglycerides is observed during starvation due to a decrease in triglyceride storage in adipose tissue ($TGL_{AP}$). The levels of free fatty acids also significantly increase during starvation, as expected based on prior research on the effects of starvation \cite{unger_eisentraut_madison_1963, yaffe_1980A}.


\subsection{Fluxes inside the cells}

\begin{figure}[tb]
    \centering
    \includegraphics[width=0.9\columnwidth]{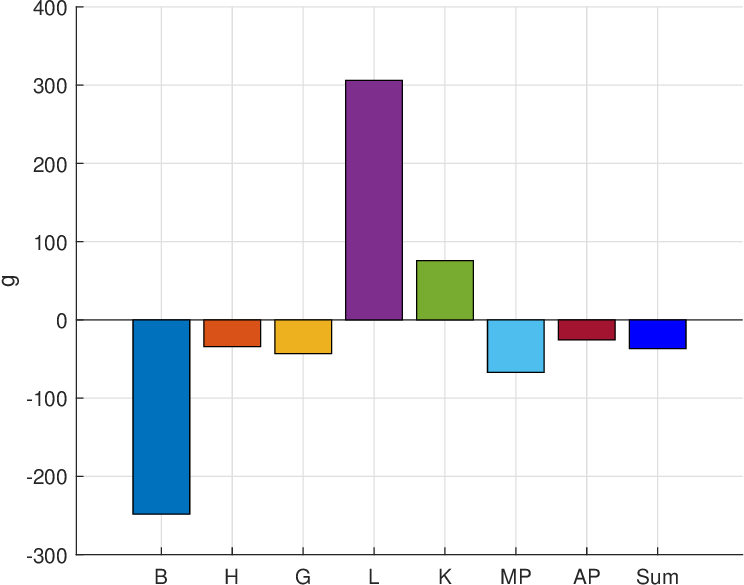}
    \caption{Sum of the glucose fluxes in every organ after an initial meal and accompanying fasting for 72 hours. The organs are the brain (B), the heart and lungs (H), the gut (G), the liver (L), the kidneys (K), the muscle tissue (MP) and the adipose tissue (AP). The total sum of all the glucose fluxes is the rightmost column denoted 'Sum'.}
    \label{fig:glucose_fluxa}
\end{figure}

\begin{figure}[tb]
    \centering
    \includegraphics[width=0.9\columnwidth]{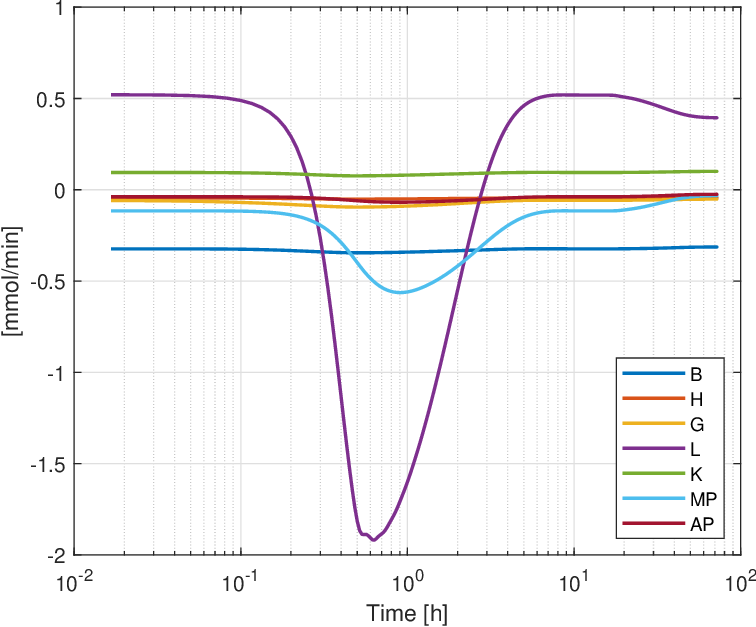}
    \caption[short]{The glucose fluxes in every organ after an initial meal and accompanying fasting for 72 hours. It shows the fluxes over time with a logarithmic scale on the x-axis.}
    \label{fig:glucose_fluxb}
\end{figure}

While it is important to note that these simulation results may not accurately reflect all metabolite levels, we can still analyze the flux dynamics. Fig. \ref{fig:glucose_fluxa} reveals that the brain is a major glucose consumer and the liver is a major glucose exporter, as expected from previous research \cite{miesfeld_mcevoy_2017}. The "Sum" column on the right shows the net flux of glucose in all organs, reflecting the overall consumption of a 60 g glucose meal. Fig. \ref{fig:glucose_fluxb} depicts which organs consume glucose over time. Initially, there is a sharp drop in glucose flux in the liver and the muscle tissue as food is ingested and blood glucose levels are high, which stimulates insulin release. As insulin levels increase, glucose uptake also increases (as shown in Table \ref{tab:insulin_glucagon_reactions}). As blood glucose levels decrease, the liver produces glucose to maintain homeostasis and its flux increases to a positive value.

\subsection{Regular food intake}
We simulate a 25 year old male at rest for 72 hours, with 4 meals a day, that has a macronutrient distribution of 60 grams of carbohydrate, 24 grams of protein and 16 grams of fat.



\begin{figure}[tb]
    \includegraphics[width=\columnwidth]{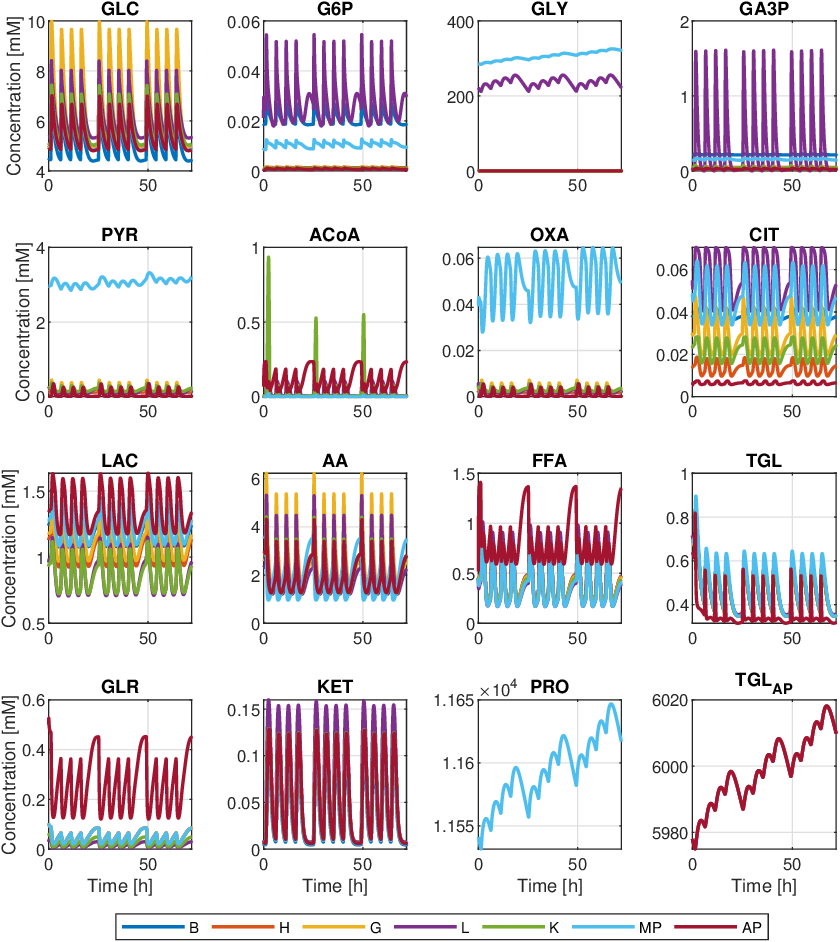}
    \caption{The metabolite concentrations of 
    glucose ($GLC$),
    glucose-6-phosphate ($G6P$),
    glycogen storage ($GLY$),
    glyceraldehyde-3-phosphate ($GA3P$),
    pyruvate ($PYR$),
    acetylcoenzyme A ($ACoA$),
    oxalate ($OXA$),
    citrate ($CIT$),
    lactate ($LAC$),
    amino acids ($AA$),
    free fatty acids ($FFA$),
    triglycerides ($TGL$),
    glycerol ($GLR$),
    ketone bodies ($KET$),
    protein storage ($PRO$),
    lipid storage ($TGL_{AP}$) in all organs after regular meals four times a day with five hours in between and then followed by nine hours of fasting during the night. Simulated over 3 days. Meals consist of 60 g glucose, 24 g protein and 16 g fat.}
    \label{fig:food_plots_bends_cars}
\end{figure}

In Fig. \ref{fig:food_plots_bends_cars} we plot all metabolite concentrations for all organs. Each spike in glucose corresponds to an intake of food. It can be seen that eating four times a day maintains a favourable glucose concentration albeit with some spikes. During the 9-hour break between simulated dinner and breakfast, it can be seen that $TGL_{AP}$ and $PRO$ acts as energy-storage and is broken down into other metabolites. Glycogen is important to maintain the glucose concentration over a short time span without a meal intake. During this regular food intake, there is only small variations in the glycogen concentration. As such, the system is able to be kept at a relatively equilibrious state.

\subsection{Simulation of intermittent fasting}
\begin{figure}[tb]
    \centering
    \includegraphics[width=\columnwidth]{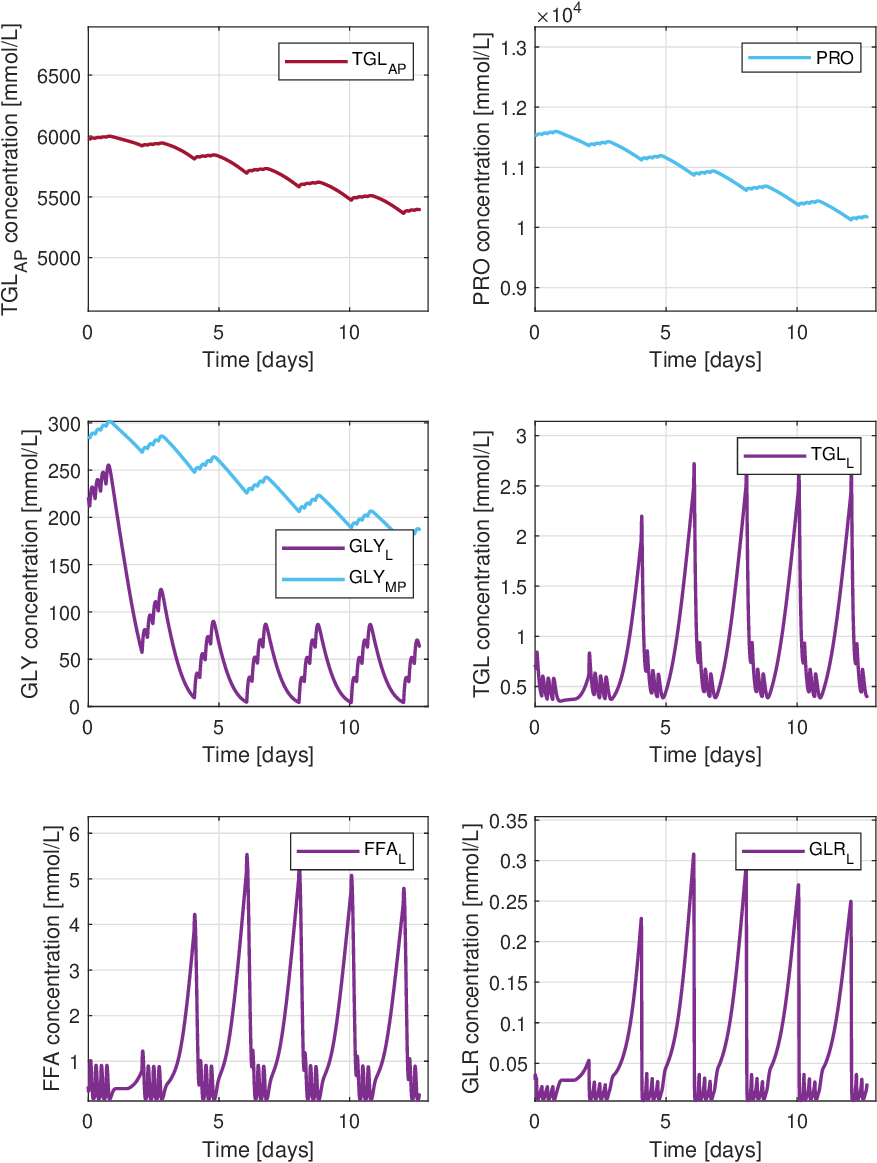}
    \caption{The metabolite concentrations of the fat storage in the adipose tissue ($TGL_{AP}$), the protein storage in the muscle tissue ($PRO$), the glycogen storage in the liver ($GLY_L$) and in the muscle tissue ($GLY_{MP}$), the triglycerides in the liver ($TGL_L$), the free fatty acids in the liver ($FFA_L$) and the glycerol in the liver ($GLR_L$). The simulation is run for 13 days with intermittent fasting every other day.}
    \label{fig:intermittent_fasting}
\end{figure}

Fig. \ref{fig:intermittent_fasting} displays a 13-day simulation with regular food intake every other day, leading to 33 hours of intermittent fasting. A reduction in lipid droplet concentration is observed as they are metabolized for energy. This is due to the shift in energy balance from equal calorie intake and consumption to a 50\% decrease in calorie intake. Protein storage in muscle tissue is also broken down as food becomes scarce. Within a day, a sharp decrease in glycogen concentration after the final meal is noted. Glycogen serves as an initial energy storage, but gets quickly depleted in the liver, while its depletion in muscle tissue is slower due to the simulated patient being at rest. Lipid droplets are transformed into FFA and GLR and used in different compartments. Large spikes occur when the body enters starvation after 18 hours without food (refer to Fig. \ref{fig:Liver_metabolites}). The simulation indicates that this diet results in weight loss as $TGL_{AP}$ (body fat) levels decrease, but also causes muscle protein loss. Intermittent fasting simulation allows for the demonstration of both regular meal effects and fasting effects.

\section{Discussion}
The cell membrane's diffusion is assumed to occur instantaneously because it equilibrates with blood vessels at a much faster rate than the timescale of the model. This assumption allows for the inclusion of organs as a single compartment. If transporters and diffusion were incorporated, the compartments would likely be divided into intracellular and extracellular to explicitly model the transporters. It is assumed that the concentration is uniform across the entire compartment. Each organ is modeled as a well-mixed tank, meaning that every metabolite is evenly distributed. Enzymatic reactions in each cell and organ occur at a fast rate of $10^{-3} \text{ to } 10^0$ seconds \cite{yasemi_jolicoeur_2021}, but for the purpose of simulating the model for several days, it is assumed that the enzymatic reactions happen at a uniform rate across the spatial organs, enabling a simplification to Michaelis-Menten kinetics.

By incorporating more metabolic inputs, this model could provide a clearer picture of an individual's dynamic changes in metabolite concentrations. The model can either be generalized or customized to a person's specific parameters, enabling personalized modeling. The current design allows for simulating enzymatic defects to depict various diseases, but validation is necessary for this purpose.

\section{Conclusion}
\label{sec:conclusion}
By adopting a systematic approach to model metabolic networks, we developed a model capable of simulating the intricate human metabolism. The approach makes it easy to extend the system through changes in the stoichiometric matrix that align with the desired chemistry and modeling objective. The core of the model lies in the parameters and reaction kinetics used in the production rate vector $R_k$. Our model, which involves 16 metabolites and 2 hormones in 7 organs, results in 126 differential equations that can be written as 7 overall differential equations for each organ, as the entire reaction network is included in $R_k$. This modeling approach can readily be expanded to incorporate larger networks of metabolic reactions within the body. The aim of the modeling was to create a foundation for a physiological whole-body model that integrates cellular metabolic processes, allowing qualitative knowledge to be used in a quantitative manner and, with proper testing, utilized for in silico trials.

\printbibliography

\onecolumn
\appendix
\setcounter{equation}{0}
\setcounter{table}{0}
\label{sec:appendix}
\renewcommand{\theequation}{\thesection.\arabic{equation}}
\renewcommand{\thetable}{\thesection.\Roman{table}}

\subsection{Model equations}

\begin{table}[H]
\centering
\resizebox{0.7\textwidth}{!}{%
\begin{tabular}{rllrl}
{  \textbf{Variables}} &                 &  & {  \textbf{Subscript}} &               \\
\textbf{C:}    & Metabolite Concentration ($\frac{mmol}{L}$)    &  & \textbf{B:}               & Brain  \\
\textbf{V:}               & Volume ($L$) &  & {  \textbf{G:}}         & Gut \\
\textbf{Q:}    & Vascular plasma flow rate ($\frac{L}{min}$)    &  &\textbf{H:}               & Heart and lungs \\
\textbf{t:}               & Time ($min$)    &  & \textbf{L:}               & Liver \\
\textbf{V$_{max}$:} & Maximum velocity ($\frac{mmol}{min}$)    &  & \textbf{K:}               & Kidney        \\
\textbf{K$_m$:}  & Limiting velocity ($\frac{mmol}{L}$)               &  & \textbf{AP:}              & Adipose   \\
\textbf{M:}  & Circulating metabolites  &  &  \textbf{MP:}            & Muscle \\
{  \textbf{Other}} &                 &  & {  \textbf{}}           &               \\
\textbf{R:}                  & Production rates             &  & {  \textbf{}}       
& \\
\textbf{r$_{oga}$:}                  & Nutrient uptake            &  & {  \textbf{}}           &                         
\end{tabular}%
}
\end{table}

\textbf{Brain:}
\begin{equation}
    V_{B}  \frac{dC_{B}}{dt} = Q_{B} M (C_H - C_B) + R_B V_B
\end{equation}

\textbf{Heart:}
\begin{equation}
    V_{H}  \frac{dC_{H}}{dt} = M (Q_{B}  C_B + Q_{L} C_L + Q_{K} C_K  + Q_{MP} C_{MP} + Q_{AP} C_{AP} - Q_{H} C_H)   + R_H V_H
\end{equation}

\textbf{Gut:}
\begin{equation}
    V_{G}  \frac{dC_{G}}{dt} = Q_{G} M (C_H - C_G) + R_G V_G + G_{r_{oga}}
\end{equation}

\textbf{Liver:}
\begin{equation}
    V_{L}  \frac{dC_{L}}{dt} = M (Q_{A} C_H + Q_{G} C_G - Q_{L} C_L) + R_L V_L
\end{equation}

\textbf{Kidney:}
\begin{equation}
    V_{K}  \frac{dC_{K}}{dt} = Q_{K} M (C_H - C_K) + R_K V_K
\end{equation}

\textbf{Muscle tissue:}
\begin{equation}
    V_{MP}  \frac{dC_{MP}}{dt} = Q_{MP} M (C_H - C_{MP}) + R_{ MP}V_{MP} + MP_{r_{oga}} 
\end{equation}

\textbf{Adipose tissue:}
\begin{equation}
    V_{AP}  \frac{dC_{AP}}{dt} = Q_{AP} M (C_H - C_{AP}) + R_{ AP}V_{AP} + AP_{r_{oga}} 
\end{equation} \\


The parameter $r_k , k \in {B,H,G,L,K,MP,AP}$ is a 31 long vector which describes the Michaelis-Menten kinetic for each reaction in each organ. The non-zero entries in $r_k$ follows the summary provided by Table \ref{tab:summary_pathways} and the universal reactions in Fig. \ref{fig:metabolic_map}. \\

\begin{equation}
    r_{k,j} = V_{max,k,j} \frac{[C_{k_i}]}{K_{m,k,j}+[C_{k_i}]} 
\end{equation}

\begin{align*}
k \in K: &\hspace{1mm} \{B,H,G,L,K,MP,AP\} \\
i \in I: &\hspace{1mm} \{GLC,G6P,GLY,GA3P,PYR,ACoA,OXA,CIT, \\
        & LAC,AA,FFA,TGL,GLR,KET,PRO,TGL_{AP}\}     \\
j \in J: &\hspace{1mm}
\{GLC\rightarrow G6P, G6P\rightarrow GLC \ldots TGL_{AP} \rightarrow 3FFA + GLR\} 
\end{align*}

The parameter M is a distribution matrix and only contains zeros and ones in the diagonal. It makes sure that only the circulating metabolites are used in the mass balance part of the equations. The parameter $r_{oga}$ connects the SIMO sub model with the main model and describes the uptake of nutrients after ingestion of food.

\onecolumn
\subsection{Kinetic equations in tissue k}

\tablefirsthead{%
    \toprule     \textbf{Reaction}&  
    &
    \textbf{} \\ 
    \midrule}

\tablehead{%
    \toprule \textbf{Reaction}& 
    & 
    \textbf{} \\ 
    \midrule}

\tabletail{%
    \midrule \multicolumn{3}{r}{{Continued on next page}} \\ 
    \midrule}

\tablelasttail{%
    \\    \midrule
    \multicolumn{3}{r}{{Concluded}} \\ 
    \bottomrule}
\topcaption{Reactions}
\label{tab:appendix_reaction}
\setlength\extrarowheight{3pt}
\begin{supertabular}{ll@{\hspace{-3mm}}l@{\hspace{0mm}}}

\textbf{1. Glycolysis 1} & & GLC $\rightarrow$ G6P \\ \shrinkheight{50pt}
& &  \large{$r_{k,GLC\rightarrow G6P} = V_{max,k,GLC\rightarrow G6P}\frac{ C_{k,GLC}}{K_{m,k,GLC\rightarrow G6P} + C_{k,GLC}}$} \\
\textbf{2. Gluconeogenesis 3} &  & G6P $\rightarrow$ G6P \\
& &   \large{$r_{k,G6P\rightarrow GLC} = V_{max,k,G6P\rightarrow GLC}\frac{ C_{k,G6P}}{K_{m,k,G6P\rightarrow GLC} + C_{k,G6P}}$} \\
\textbf{3. Glycolysis 2} & &  G6P $\rightarrow$ 2 GA3P \\
& &  \large{$r_{k,G6P\rightarrow GA3P} = V_{max,k,G6P\rightarrow GA3P}\frac{ C_{k,G6P}}{K_{m,k,G6P\rightarrow GA3P} + C_{k,G6P}}$} \\
\textbf{4. Gluconeogenesis 2}  & & 2 GA3P $\rightarrow$ G6P \\
& &  \large{$r_{k,GA3P\rightarrow G6P} = V_{max,k,GA3P\rightarrow G6P}\frac{ C_{k,GA3P}}{K_{m,k,GA3P\rightarrow G6P} + C_{k,GA3P}}$} \\
\textbf{5. Glycogenesis} & &  G6P $\rightarrow$ GLY \\
& &  \large{$r_{k,G6P\rightarrow GLY} = V_{max,k,G6P\rightarrow GLY}\frac{ C_{k,G6P}}{K_{m,k,G6P\rightarrow GLY} + C_{k,G6P}}$} \\
\textbf{6. Glycogenolysis} & & GLY $\rightarrow$ G6P \\
& &  \large{$r_{k,GLY\rightarrow G6P} = V_{max,k,GLY\rightarrow G6P}\frac{ C_{k,GLY}}{K_{m,k,GLY\rightarrow G6P} + C_{k,GLY}}$} \\
\textbf{7. Glycolysis 3} & & GA3P $\rightarrow$ PYR \\
& &  \large{$r_{k,GA3P\rightarrow PYR} = V_{max,k,GA3P\rightarrow PYR}\frac{ C_{k,GA3P}}{K_{m,k,GA3P\rightarrow PYR} + C_{k,GA3P}}$} \\
\textbf{8. Gluconeogenesis 1}  & & PYR $\rightarrow$ GA3P \\
& &  \large{$r_{k,PYR\rightarrow GA3P} = V_{max,k,PYR\rightarrow GA3P}\frac{ C_{k,PYR}}{K_{m,k,PYR\rightarrow GA3P} + C_{k,PYR}}$} \\
\textbf{9. Pyruvate fermentation}  & & PYR $\rightarrow$ LAC \\
& &  \large{$r_{k,PYR\rightarrow LAC} = V_{max,k,PYR\rightarrow LAC}\frac{ C_{k,PYR}}{K_{m,k,PYR\rightarrow LAC} + C_{k,PYR}}$} \\
\textbf{10. Lactate utilization} & & LAC $\rightarrow$ PYR \\
& &  \large{$r_{k,LAC\rightarrow PYR} = V_{max,k,LAC\rightarrow PYR}\frac{ C_{k,LAC}}{K_{m,k,LAC\rightarrow PYR} + C_{k,LAC}}$} \\
\textbf{11. Amino acid formation}  & & PYR $\rightarrow$ AA \\
& &  \large{$r_{k,PYR\rightarrow AA} = V_{max,k,PYR\rightarrow AA}\frac{ C_{k,PYR}}{K_{m,k,PYR\rightarrow AA} + C_{k,PYR}}$} \\
\textbf{12. Amino acid utilization}  & & AA $\rightarrow$ PYR \\
& &  \large{$r_{k,AA\rightarrow PYR} = V_{max,k,AA\rightarrow PYR}\frac{ C_{k,AA}}{K_{m,k,AA\rightarrow PYR} + C_{k,AA}}$} \\
\textbf{13. Pyruvate oxidation} & &  PYR $\rightarrow$ ACoA \\
& &  \large{$r_{k,PYR\rightarrow ACoA} = V_{max,k,PYR\rightarrow ACoA}\frac{ C_{k,PYR}}{K_{m,k,PYR\rightarrow ACoA} + C_{k,PYR}}$} \\
\textbf{14. Citrate formation} & &  ACoA + OXA $\rightarrow$ CIT \\
& &  \large{$r_{k,ACoA+OXA\rightarrow CIT} = V_{max,k,ACoA+OXA\rightarrow CIT}\frac{ C_{k,ACoA}C_{k,OXA}}{K_{m,k,ACoA+OXA\rightarrow CIT} + C_{k,ACoA}C_{k,OXA}}$} \\
\textbf{15. Oxaloacetate formation} & & CIT $\rightarrow$ OXA \\
& &  \large{$r_{k,CIT\rightarrow OXA} = V_{max,k,CIT\rightarrow OXA}\frac{ C_{k,CIT}}{K_{m,k,CIT\rightarrow OXA} + C_{k,CIT}}$} \\
\textbf{16. Pyruvate formation} & &  OXA $\rightarrow$ PYR \\
& &  \large{$r_{k,OXA\rightarrow PYR} = V_{max,k,OXA\rightarrow PYR}\frac{ C_{k,OXA}}{K_{m,k,OXA\rightarrow PYR} + C_{k,OXA}}$} \\
\textbf{17. Pyruvate carboxylation}  & & PYR $\rightarrow$ OXA \\
& & \large{$r_{k,PYR\rightarrow OXA} = V_{max,k,PYR\rightarrow OXA}\frac{ C_{k,PYR}}{K_{m,k,PYR\rightarrow OXA} + C_{k,PYR}}$} \\
\textbf{18. Glycerol formation} & &  GA3P $\rightarrow$ GLR \\
& & \large{$r_{k,GA3P\rightarrow GLR} = V_{max,k,GA3P\rightarrow GLR}\frac{ C_{k,GA3P}}{K_{m,k,GA3P\rightarrow GLR} + C_{k,GA3P}}$} \\

\textbf{19. Glycerol utilization} &  & GLR $\rightarrow$ GA3P \\
& &  \large{$r_{k,GLR\rightarrow GA3P} = V_{max,k,GLR\rightarrow GA3P}\frac{ C_{k,GLR}}{K_{m,k,GLR\rightarrow GA3P} + C_{k,GLR}}$} \\
\textbf{20. Esterification of fatty acids}  & & 3 FFA + GLR $\rightarrow$ TGL \\
& &  \large{$r_{k,FFA+GLR\rightarrow TGL} = V_{max,k,FFA+GLR\rightarrow TGL}\frac{ C_{k,FFA}C_{k,GLR}}{K_{m,k,FFA+GLR\rightarrow TGL} + C_{k,FFA}C_{k,GLR}}$} \\
\textbf{21. Lipolysis} & &  TGL $\rightarrow$ 3 FFA + GLR \\
& &  \large{$r_{k,TGL\rightarrow FFA+GLR} = V_{max,k,TGL\rightarrow FFA+GLR}\frac{ C_{k,TGL}}{K_{m,k,TGL\rightarrow FFA+GLR} + C_{k,TGL}}$} \\
\textbf{22. $\beta$-oxidation}  & & FFA $\rightarrow$ 7 ACoA \\
& &  \large{$r_{k,FFA\rightarrow ACoA} = V_{max,k,FFA\rightarrow ACoA}\frac{ C_{k,FFA}}{K_{m,k,FFA\rightarrow ACoA} + C_{k,FFA}}$} \\
\textbf{23. Fatty acid synthesis} & & 7 ACoA $\rightarrow$ FFA \\
& &  \large{$r_{k,ACoA\rightarrow FFA} = V_{max,k,ACoA\rightarrow FFA}\frac{ C_{k,ACoA}}{K_{m,k,ACoA\rightarrow FFA} + C_{k,ACoA}}$} \\
\textbf{24. Acetyl-CoA formation}  & & KET $\rightarrow$ ACoA \\
& &  \large{$r_{k,KET\rightarrow ACoA} = V_{max,k,KET\rightarrow ACoA}\frac{ C_{k,KET}}{K_{m,k,KET\rightarrow ACoA} + C_{k,KET}}$} \\
\textbf{25. Ketogenesis} & & ACoA $\rightarrow$ KET \\
& &  \large{$r_{k,ACoA\rightarrow KET} = V_{max,k,ACoA\rightarrow KET}\frac{ C_{k,ACoA}}{K_{m,k,ACoA\rightarrow KET} + C_{k,ACoA}}$} \\
\textbf{26. Protein anabolism}  & & AA $\rightarrow$ PRO \\
& &  \large{$r_{k,AA\rightarrow PRO} = V_{max,k,AA\rightarrow PRO}\frac{ C_{k,AA}}{K_{m,k,AA\rightarrow PRO} + C_{k,AA}}$} \\
\textbf{27. Protein catabolism}  & & PRO $\rightarrow$ AA \\
& &  \large{$r_{k,PRO\rightarrow AA} = V_{max,k,PRO\rightarrow AA}\frac{ C_{k,PRO}}{K_{m,k,PRO\rightarrow AA} + C_{k,PRO}}$} \\
\textbf{28. Lipid droplet formation}  & & 3 FFA + GLR $\rightarrow$ TGL$_{AP}$ \\
& & \normalsize{$r_{k,FFA+GLR\rightarrow TGL_{AP}} = V_{max,k,FFA+GLR\rightarrow TGL_{AP}}\frac{ C_{k,FFA}C_{k,GLR}}{K_{m,k,FFA+GLR\rightarrow TGL_{AP}} + C_{k,FFA}C_{k,GLR}}$} \\
\textbf{29. Lipid droplet hydrolysis}  & & TGL$_{AP}$ $\rightarrow$ 3 FFA + GLR \\
& & \large{$r_{k,TGL_{AP}\rightarrow FFA+GLR} = V_{max,k,TGL_{AP}\rightarrow FFA+GLR}\frac{ C_{k,TGL_{AP}}}{K_{m,k,TGL_{AP}\rightarrow FFA+GLR} + C_{k,TGL_{AP}}}$} \\
\textbf{30. Insulin reaction} & & $\rightarrow$ INS $\rightarrow$ \\ 
& & \large{$r_{k,INS} = r_{k,INS_{production}} - r_{k,INS_{clearance}}$} \\
\textbf{31. Glucagon reaction} & & $\rightarrow$ GLU $\rightarrow$ \\ 
& & \large{$r_{k,GLU} = r_{k,GLU_{production}} - r_{k,GLU_{clearance}}$} \\

\end{supertabular}%

\renewcommand{\arraystretch}{1}

\begin{table}[H]
\huge
\centering
\caption{Stoichiometric matrix. Colors match Fig. \ref{fig:metabolic_map}}
\resizebox{\textwidth}{!}{%
\begin{tabular}{c|c|c|c|c|c|c|c|c|c|c|c|c|c|c|c|c|c|c|}
\cline{2-19}
                                                      & GLC & G6P & GLY & GA3P & PYR & ACoA & OXA & CIT & LAC & AA & FFA & TGL & GLR & KET & PRO & TGL$_{AP}$ & INS & GLU \\ \hline
\rowcolor{BluePP} \multicolumn{1}{|l|}{$GLC \xrightarrow[]{} G6P$}      & -1  & 1   & 0   & 0    & 0   & 0    & 0   & 0   & 0   & 0  & 0   & 0   & 0   & 0   & 0 & 0 & 0 & 0  \\ \hline
\rowcolor{BluePP} \multicolumn{1}{|l|}{$G6P \xrightarrow[]{} GLC$}      & 1   & -1  & 0   & 0    & 0   & 0    & 0   & 0   & 0   & 0  & 0   & 0   & 0   & 0   & 0 & 0 & 0 & 0  \\ \hline
\rowcolor{BluePP} \multicolumn{1}{|l|}{$G6P \xrightarrow[]{} GA3P$}     & 0   & -1  & 0   & 2    & 0   & 0    & 0   & 0   & 0   & 0  & 0   & 0   & 0   & 0   & 0 & 0 & 0 & 0  \\ \hline
\rowcolor{BluePP} \multicolumn{1}{|l|}{$GA3P \xrightarrow[]{} G6P$}     & 0   & 1   & 0   & -2   & 0   & 0    & 0   & 0   & 0   & 0  & 0   & 0   & 0   & 0   & 0 & 0 & 0 & 0  \\ \hline
\rowcolor{BluePP} \multicolumn{1}{|l|}{$G6P \xrightarrow[]{} GLY$}      & 0   & -1  & 1   & 0    & 0   & 0    & 0   & 0   & 0   & 0  & 0   & 0   & 0   & 0   & 0 & 0 & 0 & 0  \\ \hline
\rowcolor{BluePP} \multicolumn{1}{|l|}{$GLY\xrightarrow[]{} G6P$}       & 0   & 1   & -1  & 0    & 0   & 0    & 0   & 0   & 0   & 0  & 0   & 0   & 0   & 0   & 0 & 0 & 0 & 0  \\ \hline
\rowcolor{BluePP} \multicolumn{1}{|l|}{$GA3P \xrightarrow[]{} PYR$}     & 0   & 0   & 0   & -1   & 1   & 0    & 0   & 0   & 0   & 0  & 0   & 0   & 0   & 0   & 0 & 0 & 0 & 0  \\ \hline
\rowcolor{BluePP} \multicolumn{1}{|l|}{$PYR \xrightarrow[]{} GA3P$}     & 0   & 0   & 0   & 1    & -1  & 0    & 0   & 0   & 0   & 0  & 0   & 0   & 0   & 0   & 0 & 0 & 0 & 0  \\ \hline
\rowcolor{RedPP} \multicolumn{1}{|l|}{$PYR \xrightarrow[]{} LAC$}      & 0   & 0   & 0   & 0    & -1  & 0    & 0   & 0   & 1   & 0  & 0   & 0   & 0   & 0   & 0 & 0 & 0 & 0  \\ \hline
\rowcolor{RedPP} \multicolumn{1}{|l|}{$LAC \xrightarrow[]{} PYR$}      & 0   & 0   & 0   & 0    & 1   & 0    & 0   & 0   & -1  & 0  & 0   & 0   & 0   & 0   & 0 & 0 & 0 & 0  \\ \hline
\rowcolor{PurplePP} \multicolumn{1}{|l|}{$PYR \xrightarrow[]{} AA$}       & 0   & 0   & 0   & 0    & -1  & 0    & 0   & 0   & 0   & 1  & 0   & 0   & 0   & 0   & 0 & 0 & 0 & 0  \\ \hline
\rowcolor{PurplePP} \multicolumn{1}{|l|}{$AA \xrightarrow[]{} PYR$}       & 0   & 0   & 0   & 0    & 1   & 0    & 0   & 0   & 0   & -1 & 0   & 0   & 0   & 0   & 0 & 0 & 0 & 0  \\ \hline
\rowcolor{OrangePP} \multicolumn{1}{|l|}{$PYR \xrightarrow[]{} ACoA$}     & 0   & 0   & 0   & 0    & -1  & 1    & 0   & 0   & 0   & 0  & 0   & 0   & 0   & 0   & 0 & 0 & 0 & 0  \\ \hline
\rowcolor{OrangePP} \multicolumn{1}{|l|}{$ACoA,OXA \xrightarrow[]{} CIT$} & 0   & 0   & 0   & 0    & 0   & -1   & -1  & 1   & 0   & 0  & 0   & 0   & 0   & 0   & 0 & 0 & 0 & 0  \\ \hline
\rowcolor{OrangePP} \multicolumn{1}{|l|}{$CIT \xrightarrow[]{} OXA$}      & 0   & 0   & 0   & 0    & 0   & 0    & 1   & -1  & 0   & 0  & 0   & 0   & 0   & 0   & 0 & 0  & 0 & 0 \\ \hline
\rowcolor{OrangePP} \multicolumn{1}{|l|}{$OXA \xrightarrow[]{} PYR$}      & 0   & 0   & 0   & 0    & 1   & 0    & -1  & 0   & 0   & 0  & 0   & 0   & 0   & 0   & 0 & 0 & 0 & 0  \\ \hline
\rowcolor{OrangePP} \multicolumn{1}{|l|}{$PYR \xrightarrow[]{} OXA$}      & 0   & 0   & 0   & 0    & -1  & 0    & 1   & 0   & 0   & 0  & 0   & 0   & 0   & 0   & 0 & 0  & 0 & 0 \\ \hline
\rowcolor{YellowPP} \multicolumn{1}{|l|}{$GA3P \xrightarrow[]{} GLR$}     & 0   & 0   & 0   & -1   & 0   & 0    & 0   & 0   & 0   & 0  & 0   & 0   & 1   & 0   & 0 & 0  & 0 & 0 \\ \hline
\rowcolor{YellowPP} \multicolumn{1}{|l|}{$GLR \xrightarrow[]{} GA3P$}     & 0   & 0   & 0   & 1    & 0   & 0    & 0   & 0   & 0   & 0  & 0   & 0   & -1  & 0   & 0 & 0  & 0 & 0 \\ \hline
\rowcolor{YellowPP} \multicolumn{1}{|l|}{$GLR,FFA \xrightarrow[]{} TGL$}  & 0   & 0   & 0   & 0    & 0   & 0    & 0   & 0   & 0   & 0  & -3  & 1   & -1  & 0   & 0 & 0  & 0 & 0 \\ \hline
\rowcolor{YellowPP} \multicolumn{1}{|l|}{$TGL \xrightarrow[]{} GLR,FFA$}  & 0   & 0   & 0   & 0    & 0   & 0    & 0   & 0   & 0   & 0  & 3   & -1  & 1   & 0   & 0 & 0  & 0 & 0 \\ \hline
\rowcolor{YellowPP} \multicolumn{1}{|l|}{$FFA \xrightarrow[]{} ACoA$}     & 0   & 0   & 0   & 0    & 0   & 7    & 0   & 0   & 0   & 0  & -1  & 0   & 0   & 0   & 0 & 0 & 0 & 0  \\ \hline
\rowcolor{YellowPP} \multicolumn{1}{|l|}{$ACoA \xrightarrow[]{} FFA$}     & 0   & 0   & 0   & 0    & 0   & -7   & 0   & 0   & 0   & 0  & 1   & 0   & 0   & 0   & 0 & 0  & 0 & 0 \\ \hline
\rowcolor{GreenPP} \multicolumn{1}{|l|}{$KET \xrightarrow[]{} ACoA$}     & 0   & 0   & 0   & 0    & 0   & 1    & 0   & 0   & 0   & 0  & 0   & 0   & 0   & -1  & 0 & 0  & 0 & 0 \\ \hline
\rowcolor{GreenPP} \multicolumn{1}{|l|}{$ACoA \xrightarrow[]{} KET$}     & 0   & 0   & 0   & 0    & 0   & -1   & 0   & 0   & 0   & 0  & 0   & 0   & 0   & 1   & 0 & 0  & 0 & 0 \\ \hline
\rowcolor{PurplePP} \multicolumn{1}{|l|}{$AA \xrightarrow[]{} PRO$}       & 0   & 0   & 0   & 0    & 0   & 0    & 0   & 0   & 0   & -1 & 0   & 0   & 0   & 0   & 1 & 0  & 0 & 0 \\ \hline
\rowcolor{PurplePP} \multicolumn{1}{|l|}{$PRO \xrightarrow[]{} AA$}       & 0   & 0   & 0   & 0    & 0   & 0    & 0   & 0   & 0   & 1  & 0   & 0   & 0   & 0   & -1 & 0 & 0 & 0 \\ \hline
\rowcolor{YellowPP} \multicolumn{1}{|l|}{$GLR,FFA \xrightarrow[]{} TGL_{AP}$}  & 0   & 0   & 0   & 0    & 0   & 0    & 0   & 0   & 0   & 0  & -3  & 0   & -1  & 0   & 0 & 1  & 0 & 0 \\ \hline
\rowcolor{YellowPP} \multicolumn{1}{|l|}{$TGL_{AP} \xrightarrow[]{} GLR,FFA$}  & 0   & 0   & 0   & 0    & 0   & 0    & 0   & 0   & 0   & 0  & 3   & 0  & 1   & 0   & 0  & -1  & 0 & 0 \\ \hline
\multicolumn{1}{|l|}{$\rightarrow INS \rightarrow$}  & 0   & 0   & 0   & 0    & 0   & 0    & 0   & 0   & 0   & 0  & 0  & 0   & 0  & 0   & 0 & 0  & 1 & 0 \\ \hline
\multicolumn{1}{|l|}{$\rightarrow GLU \rightarrow$}  & 0   & 0   & 0   & 0    & 0   & 0    & 0   & 0   & 0   & 0  & 0   & 0  & 0   & 0   & 0  & 0  & 0 & 1 \\ \hline
\end{tabular}%
}

\label{tab:stoichiometric_matrix}
\end{table}

\subsection{Parameter values and initial conditions}

Matlab code can be found on 

\url{https://github.com/PeterCkbs/A-whole-body-mathematical-model-for-the-metabolism-in-man}

\newpage

\twocolumn

\tablefirsthead{\toprule \textbf{Parameter} & \textbf{Value} & \textbf{Unit} \\ \midrule}
\tablehead{\toprule \textbf{Parameter} & \textbf{Value} & \textbf{Unit} \\ \midrule}

\tabletail{%
\midrule \multicolumn{3}{r}{{Continued on next column}} \\ \midrule}
\tablelasttail{%
\\\midrule
\multicolumn{3}{r}{{Concluded}} \\ \bottomrule}

\begin{supertabular}{lll}
\textbf{Flow-rates $\&$ volume} \\ \hline
$Q_B$         & 0.59      & $\frac{L}{min}$   \\ \hline
$Q_H$         & 4.37      & $\frac{L}{min}$   \\ \hline
$Q_A$         & 0.25      & $\frac{L}{min}$   \\ \hline
$Q_L$         & 1.26      & $\frac{L}{min}$   \\ \hline
$Q_G$         & 1.01      & $\frac{L}{min}$   \\ \hline
$Q_K$         & 1.01      & $\frac{L}{min}$   \\ \hline
$Q_{MP}$      & 1.223     & $\frac{L}{min}$   \\ \hline
$Q_{AP}$      & 0.287     & $\frac{L}{min}$   \\ \hline

$V_{B}$       & 0.8       & $L$       \\ \hline
$V_H$         & 1.38      & $L$       \\ \hline
$V_L$         & 2.51      & $L$       \\ \hline
$V_G$         & 1.12      & $L$       \\ \hline
$V_K$         & 0.66      & $L$       \\ \hline
$V_{AP}$      & 1.95      & $L$       \\ \hline
$V_{MP}$      & 5.84      & $L$       \\ \hline

\textbf{Brain reaction rates} \\ \hline

$V_{max, B, GLC \xrightarrow[]{} G6P}$ & 0.4954 & $\frac{mM}{min}$       \\ \hline
$K_{m, B,GLC \xrightarrow[]{} G6P}$ & 1 & $mM$   \\ \hline

$V_{max, B,G6P \xrightarrow[]{} GA3P}$ & 0.4954 & $\frac{mM}{min}$       \\ \hline
$K_{m, B,G6P \xrightarrow[]{} GA3P}$ & 0.0042 & $mM$   \\ \hline

$V_{max, B,GA3P \xrightarrow[]{} PYR}$ & 7.838 & $\frac{mM}{min}$       \\ \hline
$K_{m, B,GA3P \xrightarrow[]{} PYR}$ & 1.86 & $mM$   \\ \hline

$V_{max, B,PYR \xrightarrow[]{} LAC}$ & 0.6964 & $\frac{mM}{min}$       \\ \hline
$K_{m, B,PYR \xrightarrow[]{} LAC}$ & 0.065 & $mM$   \\ \hline

$V_{max, B,LAC \xrightarrow[]{} PYR}$ & 1.807 & $\frac{mM}{min}$       \\ \hline
$K_{m, B,LAC \xrightarrow[]{} PYR}$ & 4.507 & $mM$   \\ \hline

$V_{max, B,PYR \xrightarrow[]{} ACoA}$ & 1.738 & $\frac{mM}{min}$       \\ \hline
$K_{m, B,PYR \xrightarrow[]{} ACoA}$ & 0.187 & $mM$   \\ \hline

$V_{max, B,ACoA,OXA \xrightarrow[]{} CIT}$ & 1.043 & $\frac{mM}{min}$       \\ \hline
$K_{m, B,ACoA,OXA \xrightarrow[]{} CIT}$ & 4.5e-6  & $mM$   \\ \hline

$V_{max, B,CIT \xrightarrow[]{} OXA}$ & 49.18 & $\frac{mM}{min}$       \\ \hline
$K_{m, B,CIT \xrightarrow[]{} OXA}$ & 2.51 & $mM$   \\ \hline

$V_{max, B,OXA \xrightarrow[]{} PYR}$ & 0.0125 & $\frac{mM}{min}$       \\ \hline
$K_{m, B,OXA \xrightarrow[]{} PYR}$ & 0.003 & $mM$   \\ \hline

$V_{max, B,PYR \xrightarrow[]{} OXA}$ & 0.0125 & $\frac{mM}{min}$       \\ \hline
$K_{m, B,PYR \xrightarrow[]{} OXA}$ & 0.187 & $mM$   \\ \hline

$V_{max, B,KET \xrightarrow[]{} ACoA}$ & 1.25 & $\frac{mM}{min}$       \\ \hline
$K_{m, B,KET \xrightarrow[]{} ACoA}$ & 5 & $mM$   \\ \hline

\textbf{Heart reaction rates} \\ \hline
$V_{max, H,GLC \xrightarrow[]{} G6P}$ & 0.09 & $\frac{mM}{min}$       \\ \hline
$K_{m, H,GLC \xrightarrow[]{} G6P}$ & 3 & $mM$   \\ 

$V_{max, H,G6P \xrightarrow[]{} GA3P}$ & 0.2377 & $\frac{mM}{min}$       \\ \hline
$K_{m, H,G6P \xrightarrow[]{} GA3P}$ & 4.2e-3 & $mM$   \\ \hline

$V_{max, H,GA3P \xrightarrow[]{} PYR}$ & 4.544 & $\frac{mM}{min}$       \\ \hline
$K_{m, H,GA3P \xrightarrow[]{} PYR}$ & 1.86 & $mM$   \\ \hline

$V_{max, H,PYR \xrightarrow[]{} LAC}$ & 0.4432 & $\frac{mM}{min}$       \\ \hline
$K_{m, H,PYR \xrightarrow[]{} LAC}$ & 0.122 & $mM$   \\ \hline

$V_{max, H,LAC \xrightarrow[]{} PYR}$ & 0.5745 & $\frac{mM}{min}$       \\ \hline
$K_{m, H,LAC \xrightarrow[]{} PYR}$ & 3.371 & $mM$   \\ \hline

$V_{max, H,PYR \xrightarrow[]{} ACoA}$ & 0.0658  & $\frac{mM}{min}$       \\ \hline
$K_{m, H,PYR \xrightarrow[]{} ACoA}$ & 0.187 & $mM$   \\ \hline

$V_{max, H,ACoA , OXA \xrightarrow[]{} CIT}$ & 0.2228 & $\frac{mM}{min}$       \\ \hline
$K_{m, H,ACoA , OXA \xrightarrow[]{} CIT}$ & 4.5e-6  & $mM$   \\ \hline

$V_{max, H,CIT \xrightarrow[]{} OXA}$ & 28.51 & $\frac{mM}{min}$       \\ \hline
$K_{m, H,CIT \xrightarrow[]{} OXA}$ & 2.51 & $mM$   \\ \hline

$V_{max, H,OXA \xrightarrow[]{} PYR}$ & 0.0072 & $\frac{mM}{min}$       \\ \hline
$K_{m, H,OXA \xrightarrow[]{} PYR}$ & 0.003 & $mM$   \\ \hline

$V_{max, H,PYR \xrightarrow[]{} OXA}$ & 0.0072 & $\frac{mM}{min}$       \\ \hline
$K_{m, H,PYR \xrightarrow[]{} OXA}$ & 0.187 & $mM$   \\ \hline

$V_{max, H,TGL \xrightarrow[]{} GLR , FFA}$ & 0.0023 & $\frac{mM}{min}$       \\ \hline
$K_{m, H,TGL \xrightarrow[]{} GLR , FFA}$ & 11.21 & $mM$   \\ \hline

$V_{max, H,FFA \xrightarrow[]{} ACoA}$ & 0.0383 & $\frac{mM}{min}$       \\ \hline
$K_{m, H,FFA \xrightarrow[]{} ACoA}$ & 0.45 & $mM$   \\ \hline

$V_{max, H,KET \xrightarrow[]{} ACoA}$ & 0.0186 & $\frac{mM}{min}$       \\ \hline
$K_{m, H,KET \xrightarrow[]{} ACoA}$ & 0.5 & $mM$   \\ \hline

\textbf{Gut reaction rates} \\ \hline
$V_{max, G,GLC \xrightarrow[]{} G6P}$ & 0.2289 & $\frac{mM}{min}$       \\ \hline
$K_{m, G,GLC \xrightarrow[]{} G6P}$ & 17 & $mM$   \\ \hline

$V_{max, G,G6P \xrightarrow[]{} GA3P}$ & 0.2929 & $\frac{mM}{min}$       \\ \hline
$K_{m, G,G6P \xrightarrow[]{} GA3P}$ & 4.2e-3 & $mM$   \\ \hline

$V_{max, G,GA3P \xrightarrow[]{} PYR}$ & 5.598 & $\frac{mM}{min}$       \\ \hline
$K_{m, G,GA3P \xrightarrow[]{} PYR}$ & 1.86 & $mM$   \\ \hline

$V_{max, G,PYR \xrightarrow[]{} LAC}$ & 0.5555 & $\frac{mM}{min}$       \\ \hline
$K_{m, G,PYR \xrightarrow[]{} LAC}$ & 0.039 & $mM$   \\ \hline

$V_{max, G,LAC \xrightarrow[]{} PYR}$ & 0.8705 & $\frac{mM}{min}$       \\ \hline
$K_{m, G,LAC \xrightarrow[]{} PYR}$ & 1.199 & $mM$   \\ \hline

$V_{max, G,AA \xrightarrow[]{} PYR}$ & 1.009 & $\frac{mM}{min}$       \\ \hline
$K_{m, G,AA \xrightarrow[]{} PYR}$ & 8.55 & $mM$   \\ \hline

$V_{max, G,PYR \xrightarrow[]{} ACoA}$ & 1.199 & $\frac{mM}{min}$       \\ \hline
$K_{m, G,PYR \xrightarrow[]{} ACoA}$ & 0.561 & $mM$   \\ \hline

$V_{max, G,ACoA , OXA \xrightarrow[]{} CIT}$ & 0.7993 & $\frac{mM}{min}$       \\ \hline
$K_{m, G,ACoA , OXA \xrightarrow[]{} CIT}$ & 4.5e-6 & $mM$   \\ \hline

$V_{max, G,CIT \xrightarrow[]{} OXA}$ & 35.13 & $\frac{mM}{min}$       \\ 
$K_{m, G,CIT \xrightarrow[]{} OXA}$ & 2.51 & $mM$   \\ \hline

$V_{max, G,OXA \xrightarrow[]{} PYR}$ & 0.0089 & $\frac{mM}{min}$       \\ \hline
$K_{m, G,OXA \xrightarrow[]{} PYR}$ & 0.003 & $mM$   \\ \hline

$V_{max, G,PYR \xrightarrow[]{} OXA}$ & 0.0089 & $\frac{mM}{min}$       \\ \hline
$K_{m, G,PYR \xrightarrow[]{} OXA}$ & 0.187 & $mM$   \\ \hline

$V_{max, G,FFA \xrightarrow[]{} ACoA}$ & 0.0571 & $\frac{mM}{min}$       \\ \hline
$K_{m, G,FFA \xrightarrow[]{} ACoA}$ & 1.35 & $mM$   \\ \hline

\textbf{Liver reaction rates} \\ \hline
$V_{max, L,GLC \xrightarrow[]{} G6P}$ & 0.1703 & $\frac{mM}{min}$       \\ \hline
$K_{m, L,GLC \xrightarrow[]{} G6P}$ & 17 & $mM$   \\ \hline

$V_{max, L,G6P \xrightarrow[]{} GLC}$ & 0.2478 & $\frac{mM}{min}$       \\ \hline
$K_{m, L,G6P \xrightarrow[]{} GLC}$ & 1-e5 & $mM$   \\ \hline

$V_{max, L,G6P \xrightarrow[]{} GA3P}$ & 0.1307 & $\frac{mM}{min}$       \\ \hline
$K_{m, L,G6P \xrightarrow[]{} GA3P}$ & 0.258 & $mM$   \\ \hline

$V_{max, L,GA3P \xrightarrow[]{} G6P}$ & 0.0929 & $\frac{mM}{min}$       \\ \hline
$K_{m, L,GA3P \xrightarrow[]{} G6P}$ & 1e-5 & $mM$   \\ \hline

$V_{max, L,G6P \xrightarrow[]{} GLY}$ & 0.2716 & $\frac{mM}{min}$       \\ \hline
$K_{m, L,G6P \xrightarrow[]{} GLY}$ & 0.258 & $mM$   \\ \hline

$V_{max, L,GLY \xrightarrow[]{} G6P}$ & 0.31 & $\frac{mM}{min}$       \\ \hline
$K_{m, L,GLY \xrightarrow[]{} G6P}$ & 221.3 & $mM$   \\ \hline

$V_{max, L,GA3P \xrightarrow[]{} PYR}$ & 2.498 & $\frac{mM}{min}$       \\ \hline
$K_{m, L,GA3P \xrightarrow[]{} PYR}$ & 1.86 & $mM$   \\ \hline

$V_{max, L,PYR \xrightarrow[]{} GA3P}$ & 0.3073 & $\frac{mM}{min}$       \\ \hline
$K_{m, L,PYR \xrightarrow[]{} GA3P}$ & 0.187 & $mM$   \\ \hline

$V_{max, L,PYR \xrightarrow[]{} LAC}$ & 0.3565 & $\frac{mM}{min}$       \\ \hline
$K_{m, L,PYR \xrightarrow[]{} LAC}$ & 0.241 & $mM$   \\ \hline

$V_{max, L,LAC \xrightarrow[]{} PYR}$ & 0.5617 & $\frac{mM}{min}$       \\ \hline
$K_{m, L,LAC \xrightarrow[]{} PYR}$ & 1.052 & $mM$   \\ \hline

$V_{max, L,AA \xrightarrow[]{} PYR}$ & 0.4212 & $\frac{mM}{min}$       \\ \hline
$K_{m, L,AA \xrightarrow[]{} PYR}$ & 2.85 & $mM$   \\ \hline

$V_{max, L,PYR \xrightarrow[]{} ACoA}$ & 0.2733 & $\frac{mM}{min}$       \\ \hline
$K_{m, L,PYR \xrightarrow[]{} ACoA}$ & 0.187 & $mM$   \\ \hline

$V_{max, L,ACoA , OXA \xrightarrow[]{} CIT}$ & 0.433 & $\frac{mM}{min}$       \\ \hline
$K_{m, L,ACoA , OXA \xrightarrow[]{} CIT}$ & 4.5e-6 & $mM$   \\ \hline

$V_{max, L,CIT \xrightarrow[]{} OXA}$ & 15.67 & $\frac{mM}{min}$       \\ \hline
$K_{m, L,CIT \xrightarrow[]{} OXA}$ & 2.51 & $mM$   \\ \hline

$V_{max, L,OXA \xrightarrow[]{} PYR}$ & 0.004 & $\frac{mM}{min}$       \\ \hline
$K_{m, L,OXA \xrightarrow[]{} PYR}$ & 0.003 & $mM$   \\ \hline

$V_{max, L,PYR \xrightarrow[]{} OXA}$ & 0.004 & $\frac{mM}{min}$       \\ \hline
$K_{m, L,PYR \xrightarrow[]{} OXA}$ & 0.187 & $mM$   \\ \hline

$V_{max, L,GLR \xrightarrow[]{} GA3P}$ & 0.0642 & $\frac{mM}{min}$       \\ \hline
$K_{m, L,GLR \xrightarrow[]{} GA3P}$ & 0.05 & $mM$   \\ 

$V_{max, L,GLR , FFA \xrightarrow[]{} TGL}$ & 0.017 & $\frac{mM}{min}$       \\ \hline
$K_{m, L,GLR , FFA \xrightarrow[]{} TGL}$ & 0.0178 & $mM$   \\ \hline

$V_{max, L,FFA \xrightarrow[]{} ACoA}$ & 0.1195 & $\frac{mM}{min}$       \\ \hline
$K_{m, L,FFA \xrightarrow[]{} ACoA}$ & 1.385 & $mM$   \\ \hline

$V_{max, L,ACoA \xrightarrow[]{} FFA}$ & 0.1195 & $\frac{mM}{min}$       \\ \hline
$K_{m, L,ACoA \xrightarrow[]{} FFA}$ & 0.5 & $mM$   \\ \hline

$V_{max, L,ACoA \xrightarrow[]{} KET}$ & 0.3984 & $\frac{mM}{min}$       \\ \hline
$K_{m, L,ACoA \xrightarrow[]{} KET}$ & 2 & $mM$   \\ \hline

\textbf{Kidney reaction rates} \\ \hline
$V_{max, K,GLC \xrightarrow[]{} G6P}$ & 0.3885 & $\frac{mM}{min}$       \\ \hline
$K_{m, K,GLC \xrightarrow[]{} G6P}$ & 17 & $mM$   \\ \hline

$V_{max, K,G6P \xrightarrow[]{} GLC}$ & 0.2355 & $\frac{mM}{min}$       \\ \hline
$K_{m, K,G6P \xrightarrow[]{} GLC}$ & 1e-5 & $mM$   \\ \hline

$V_{max, K,G6P \xrightarrow[]{} GA3P}$ & 0.497 & $\frac{mM}{min}$       \\ \hline
$K_{m, K,G6P \xrightarrow[]{} GA3P}$ & 4.2-3 & $mM$   \\ \hline

$V_{max, K,GA3P \xrightarrow[]{} G6P}$ & 0.2355 & $\frac{mM}{min}$       \\ \hline
$K_{m, K,GA3P \xrightarrow[]{} G6P}$ & 1e-5 & $mM$   \\ \hline

$V_{max, K,GA3P \xrightarrow[]{} PYR}$ & 9.5 & $\frac{mM}{min}$       \\ \hline
$K_{m, K,GA3P \xrightarrow[]{} PYR}$ & 1.86 & $mM$   \\ \hline

$V_{max, K,PYR \xrightarrow[]{} GA3P}$ & 0.8306 & $\frac{mM}{min}$       \\ \hline
$K_{m, K,PYR \xrightarrow[]{} GA3P}$ & 0.187 & $mM$   \\ \hline

$V_{max, K,PYR \xrightarrow[]{} LAC}$ & 0.3718 & $\frac{mM}{min}$       \\ \hline
$K_{m, K,PYR \xrightarrow[]{} LAC}$ & 0.3045 & $mM$   \\ \hline

$V_{max, K,LAC \xrightarrow[]{} PYR}$ & 0.6279 & $\frac{mM}{min}$       \\ \hline
$K_{m, K,LAC \xrightarrow[]{} PYR}$ & 0.324 & $mM$   \\ \hline

$V_{max, K,AA \xrightarrow[]{} PYR}$ & 0.2485 & $\frac{mM}{min}$       \\ \hline
$K_{m, K,AA \xrightarrow[]{} PYR}$ & 2.85 & $mM$   \\ \hline

$V_{max, K,PYR \xrightarrow[]{} ACoA}$ & 0.3533 & $\frac{mM}{min}$       \\ \hline
$K_{m, K,PYR \xrightarrow[]{} ACoA}$ & 0.187 & $mM$   \\ \hline

$V_{max, K,ACoA , OXA \xrightarrow[]{} CIT}$ & 0.6655 & $\frac{mM}{min}$       \\ \hline
$K_{m, K,ACoA , OXA \xrightarrow[]{} CIT}$ & 4.5e-6 & $mM$   \\ \hline

$V_{max, K,CIT \xrightarrow[]{} OXA}$ & 59.61 & $\frac{mM}{min}$       \\ \hline
$K_{m, K,CIT \xrightarrow[]{} OXA}$ & 2.51 & $mM$   \\ \hline

$V_{max, K,OXA \xrightarrow[]{} PYR}$ & 0.0152 & $\frac{mM}{min}$       \\ \hline
$K_{m, K,OXA \xrightarrow[]{} PYR}$ & 0.003 & $mM$   \\ \hline

$V_{max, K,PYR \xrightarrow[]{} OXA}$ & 0.0152 & $\frac{mM}{min}$       \\ \hline
$K_{m, K,PYR \xrightarrow[]{} OXA}$ & 0.187 & $mM$   \\ \hline

$V_{max, K,GLR \xrightarrow[]{} GA3P}$ & 0.1115 & $\frac{mM}{min}$       \\ \hline
$K_{m, K,GLR \xrightarrow[]{} GA3P}$ & 0.05 & $mM$   \\ \hline

$V_{max, K,FFA \xrightarrow[]{} ACoA}$ & 0.1079 & $\frac{mM}{min}$       \\ \hline
$K_{m, K,FFA \xrightarrow[]{} ACoA}$ & 0.45 & $mM$   \\ \hline

\textbf{Muscle reaction rates} \\ 
$V_{max, MP,GLC \xrightarrow[]{} G6P}$ & 0.0399 & $\frac{mM}{min}$       \\ \hline
$K_{m, MP,GLC \xrightarrow[]{} G6P}$ & 5 & $mM$   \\ \hline

$V_{max, MP,G6P \xrightarrow[]{} GA3P}$ & 0.0562 & $\frac{mM}{min}$       \\ \hline
$K_{m, MP,G6P \xrightarrow[]{} GA3P}$ & 4.2e-3 & $mM$   \\ \hline

$V_{max, MP,G6P \xrightarrow[]{} GLY}$ & 0.0855 & $\frac{mM}{min}$       \\ \hline
$K_{m, MP,G6P \xrightarrow[]{} GLY}$ & 8.93e-2 & $mM$   \\ \hline

$V_{max, MP,GLY \xrightarrow[]{} G6P}$ & 0.076 & $\frac{mM}{min}$       \\ \hline
$K_{m, MP,GLY \xrightarrow[]{} G6P}$ & 571.2 & $mM$   \\ \hline

$V_{max, MP,GA3P \xrightarrow[]{} PYR}$ & 1.075 & $\frac{mM}{min}$       \\ \hline
$K_{m, MP,GA3P \xrightarrow[]{} PYR}$ & 1.86 & $mM$   \\ \hline

$V_{max, MP,PYR \xrightarrow[]{} LAC}$ & 0.1407 & $\frac{mM}{min}$       \\ \hline
$K_{m, MP,PYR \xrightarrow[]{} LAC}$ & 0.503 & $mM$   \\ \hline

$V_{max, MP,LAC \xrightarrow[]{} PYR}$ & 0.1174 & $\frac{mM}{min}$       \\ \hline
$K_{m, MP,LAC \xrightarrow[]{} PYR}$ & 0.382 & $mM$   \\ \hline

$V_{max, MP,PYR \xrightarrow[]{} AA}$ & 0.0133 & $\frac{mM}{min}$       \\ \hline
$K_{m, MP,PYR \xrightarrow[]{} AA}$ & 7.43e-4 & $mM$   \\ \hline

$V_{max, MP,PYR \xrightarrow[]{} ACoA}$ & 0.0339 & $\frac{mM}{min}$       \\ \hline
$K_{m, MP,PYR \xrightarrow[]{} ACoA}$ & 0.187 & $mM$   \\ \hline

$V_{max, MP,ACoA , OXA \xrightarrow[]{} CIT}$ & 0.1693 & $\frac{mM}{min}$       \\ \hline
$K_{m, MP,ACoA , OXA \xrightarrow[]{} CIT}$ & 4.5e-6 & $mM$   \\ \hline

$V_{max, MP,CIT \xrightarrow[]{} OXA}$ & 6.742 & $\frac{mM}{min}$       \\ \hline
$K_{m, MP,CIT \xrightarrow[]{} OXA}$ & 2.51 & $mM$   \\ \hline

$V_{max, MP,OXA \xrightarrow[]{} PYR}$ & 0.0017 & $\frac{mM}{min}$       \\ \hline
$K_{m, MP,OXA \xrightarrow[]{} PYR}$ & 0.003 & $mM$   \\ 
\hline

$V_{max, MP,PYR \xrightarrow[]{} OXA}$ & 0.0017 & $\frac{mM}{min}$       \\ \hline
$K_{m, MP,PYR \xrightarrow[]{} OXA}$ & 0.187 & $mM$   \\ \hline

$V_{max, MP,TGL \xrightarrow[]{} GLR , FFA}$ & 5.48e-4 & $\frac{mM}{min}$       \\ \hline
$K_{m, MP,TGL \xrightarrow[]{} GLR , FFA}$ & 11.21 & $mM$   \\ \hline

$V_{max, MP,FFA \xrightarrow[]{} ACoA}$ & 0.0274 & $\frac{mM}{min}$       \\ \hline
$K_{m, MP,FFA \xrightarrow[]{} ACoA}$ & 0.45 & $mM$   \\ \hline

$V_{max, MP,KET \xrightarrow[]{} ACoA}$ & 0.0113 & $\frac{mM}{min}$       \\ \hline
$K_{m, MP,KET \xrightarrow[]{} ACoA}$ & 0.5 & $mM$   \\ \hline

$V_{max, MP,AA \xrightarrow[]{} PRO}$ & 0.329 & $\frac{mM}{min}$       \\ \hline
$K_{m, MP,AA \xrightarrow[]{} PRO}$ & 11.4 & $mM$   \\ \hline

$V_{max, MP,PRO \xrightarrow[]{} AA}$ & 0.4113 & $\frac{mM}{min}$       \\ \hline
$K_{m, MP,PRO \xrightarrow[]{} AA}$ & 11541 & $mM$   \\ \hline

\textbf{Adipose reaction rates} \\ \hline
$V_{max, AP,GLC \xrightarrow[]{} G6P}$ & 0.0399 & $\frac{mM}{min}$       \\ \hline
$K_{m, AP,GLC \xrightarrow[]{} G6P}$ & 5 & $mM$   \\ \hline

$V_{max, AP,G6P \xrightarrow[]{} GA3P}$ & 01686 & $\frac{mM}{min}$       \\ \hline
$K_{m, AP,G6P \xrightarrow[]{} GA3P}$ & 4.2e-3 & $mM$   \\ \hline

$V_{max, AP,GA3P \xrightarrow[]{} PYR}$ & 3.224 & $\frac{mM}{min}$       \\ 
$K_{m, AP,GA3P \xrightarrow[]{} PYR}$ & 1.86 & $mM$   \\ \hline

$V_{max, AP,PYR \xrightarrow[]{} LAC}$ & 0.2656 & $\frac{mM}{min}$       \\ \hline
$K_{m, AP,PYR \xrightarrow[]{} LAC}$ & 2.2e-3 & $mM$   \\ \hline

$V_{max, AP,LAC \xrightarrow[]{} PYR}$ & 0.4795 & $\frac{mM}{min}$       \\ \hline
$K_{m, AP,LAC \xrightarrow[]{} PYR}$ & 1.98 & $mM$   \\ \hline

$V_{max, AP,PYR \xrightarrow[]{} ACoA}$ & 0.0439 & $\frac{mM}{min}$       \\ \hline
$K_{m, AP,PYR \xrightarrow[]{} ACoA}$ & 1.87 & $mM$   \\ \hline

$V_{max, AP,ACoA , OXA \xrightarrow[]{} CIT}$ & 0.0608 & $\frac{mM}{min}$       \\ \hline
$K_{m, AP,ACoA , OXA \xrightarrow[]{} CIT}$ & 4.5e-6 & $mM$   \\ \hline

$V_{max, AP,CIT \xrightarrow[]{} OXA}$ & 20.23 & $\frac{mM}{min}$       \\ \hline
$K_{m, AP,CIT \xrightarrow[]{} OXA}$ & 2.51 & $mM$   \\ \hline

$V_{max, AP,OXA \xrightarrow[]{} PYR}$ & 0.0051 & $\frac{mM}{min}$       \\ \hline
$K_{m, AP,OXA \xrightarrow[]{} PYR}$ & 0.003 & $mM$   \\ \hline

$V_{max, AP,PYR \xrightarrow[]{} OXA}$ & 0.0051 & $\frac{mM}{min}$       \\ \hline
$K_{m, AP,PYR \xrightarrow[]{} OXA}$ & 0.187 & $mM$   \\ \hline

$V_{max, AP,GA3P \xrightarrow[]{} GLR}$ & 0.0159 & $\frac{mM}{min}$       \\ \hline
$K_{m, AP,GA3P \xrightarrow[]{} GLR}$ & 0.016 & $mM$   \\ \hline

$V_{max, AP,TGL \xrightarrow[]{} GLR , FFA}$ & 0.9202 & $\frac{mM}{min}$       \\ \hline
$K_{m, AP,TGL \xrightarrow[]{} GLR , FFA}$ & 63.53 & $mM$   \\ \hline

$V_{max, AP,FFA \xrightarrow[]{} ACoA}$ & 0.0133 & $\frac{mM}{min}$       \\ \hline
$K_{m, AP,FFA \xrightarrow[]{} ACoA}$ & 0.45 & $mM$   \\ \hline

$V_{max, AP,ACoA \xrightarrow[]{} FFA}$ & 0.0133 & $\frac{mM}{min}$       \\ \hline
$K_{m, AP,ACoA \xrightarrow[]{} FFA}$ & 1 & $mM$   \\ \hline

$V_{max, AP,GLR, FFA \xrightarrow[]{} TGL_{AP}}$ & 0.184 & $\frac{mM}{min}$       \\ \hline
$K_{m, AP,GLR, FFA \xrightarrow[]{} TGL_{AP}}$ & 0.1368 & $mM$   \\ \hline

$V_{max, AP,TGL_{AP} \xrightarrow[]{} GLR, FFA}$ & 0.3247 & $\frac{mM}{min}$       \\ \hline
$K_{m, AP,TGL_{AP} \xrightarrow[]{} GLR, FFA}$ & 5978 & $mM$   \\ \hline

\textbf{Insulin parameters} \\ \hline
$F_{LIC}^I$     & 0.4       &  $\frac{mU}{min}$       \\ \hline
$F_{KIC}^I$     & 0.3       &  $\frac{mU}{min}$      \\ \hline
$F_{PIC}^I$     & 0.15      &  $\frac{mU}{min}$      \\  \hline

$\beta_{PIR1}^I$     & 3.27      &  -      \\ \hline
$\beta_{PIR2}^I$     & 132       & $\frac{mg}{dL}$       \\ \hline
$\beta_{PIR3}^I$     & 5.93      &  -      \\ \hline
$\beta_{PIR4}^I$     & 3.02      &  -      \\ \hline
$\beta_{PIR5}^I$     & 1.11      &  -      \\ \hline
$M_{2}^I$       & 7.47e-3      &    -    \\ \hline
$M_{2}^I$       & 9.58e-2      &    -    \\ \hline
$\alpha^I$      & 4.82e-2        & $\frac{1}{min}$       \\ \hline
$\beta^I$       & 0.931       & $\frac{1}{min}$       \\ \hline
$Q_{0}^I$       & 6.33       & U       \\ \hline
$K^I$           & 7.94e-3      &    -    \\ 
$y^I$           & 0.575       &     -   \\ \hline

\textbf{Glucagon parameters} \\ \hline
r$_{M\Gamma C}$         & 0.91      & $\frac{L}{min}$   \\ \hline

\textbf{Hormonal $\mu$ values} \\ \hline
$\mu_{L, GLC \rightarrow G6P}$            & 4.2422 & - \\ \hline
$\mu_{L, G6P \rightarrow GLY}$            & 2      & - \\ \hline
$\mu_{L, GLY \rightarrow G6P}$            & 0.5    & - \\ \hline
$\mu_{L, G6P \rightarrow GA3P}$           & 3      & - \\ \hline
$\mu_{L, GA3P \rightarrow G6P}$           & 0.5    & - \\ \hline
$\mu_{L, PYR \rightarrow ACoA}$           & 3      & - \\ \hline
$\mu_{L, ACoA \rightarrow FFA}$           & 3      & - \\ \hline
$\mu_{MP, GLC \rightarrow G6P}$           & 2.1557 & - \\ \hline
$\mu_{MP, G6P \rightarrow GLY}$           & 3      & - \\ \hline
$\mu_{MP, GLY \rightarrow G6P}$           & 0.2    & - \\ \hline
$\mu_{MP, G6P \rightarrow GA3P}$          & 0.1    & - \\ \hline
$\mu_{MP, AA \rightarrow PRO}$            & 4      & - \\ \hline
$\mu_{MP, PRO \rightarrow AA}$            & 0.25      & - \\ \hline
$\mu_{MP, PYR \rightarrow ACoA}$          & 0.5      & - \\ \hline
$\mu_{AP, GLC \rightarrow G6P}$           & 0.6352 & - \\ \hline
$\mu_{AP, TGL \rightarrow GLR, FFA}$      & 4      & - \\ 
\hline
$\mu_{AP, TGL_{AP} \rightarrow GLR, FFA}$ & 0.1    & - \\ \hline
$\mu_{AP, GLR, FFA \rightarrow TGL_{AP}}$ & 2      & - \\ \hline

\textbf{SIMO model parameters} \\ \hline
$k_{js}$        & 0.028237  & $min^{-1}$ \\ \hline
$k_{gl}$        & 0.0180942 & $min^{-1}$ \\ \hline
$k_{gj}$        & 0.0329673 & $min^{-1}$ \\ \hline
$k_{rj}$        & 0.0344046 & $min^{-1}$ \\ \hline
$k_{lr}$        & 0.0513802 & $min^{-1}$  \\ \hline
\end{supertabular}%

\onecolumn
\newpage

\begin{table}[H]
\centering
\caption{Initial values table for metabolites in the model}
\label{tab:initial_values}
\begin{tabular}{|l|l|l|l|l|l|l|l|}
\hline
\textbf{Initial values} & \textbf{Brain} & \textbf{Heart} & \textbf{Gut} & \textbf{Liver} & \textbf{Kidney} & \textbf{Muscle} & \textbf{Adipose} \\ \hline
$\text{GLC}_0$ [mM]           & 4.44     & 5         & 4.93      & 5.36      & 5.1       & 4.9       & 4.85 \\ \hline
$\text{G6P}_0$ [mM]           & 0.0187   & 0.0013    & 0.0009    & 0.0212    & 0.001     & 0.0084    & 0.0006 \\ \hline
$\text{GLY}_0$    [mM]        & 0        & 0         & 0         & 221.3     & 0         & 285.63    & 0      \\ \hline
$\text{GA3P}_0$ [mM]          & 0.214    & 0.047     & 0.035     & 0.016     & 0.047     & 0.139     & 0.018 \\ \hline
$\text{PYR}_0$    [mM]        & 0.135    & 0.132     & 0.195     & 0.22      & 0.239     & 2.96      & 0.012 \\ \hline
$\text{ACoA}_0$ [mM]          & 0.005    & 0.006     & 0.002     & 0.005     & 0.008     & 0.0004    & 0.0683 \\ \hline
$\text{OXA}_0$    [mM]        & 0.002    & 0.002     & 0.003     & 0.004     & 0.004     & 0.04      & 0.0002 \\ \hline
$\text{CIT}_0$    [mM]        & 0.0376   & 0.0143    & 0.03      & 0.055     & 0.024     & 0.048     & 0.0055 \\ \hline
$\text{LAC}_0$    [mM]        & 1.24     & 1.13      & 1.17      & 0.97      & 0.94      & 1.28      & 1.34 \\ \hline
$\text{AA}_0$ [mM]            & 2.91     & 2.91      & 2.65      & 2.32      & 2.83      & 3.59      & 2.91 \\ \hline
$\text{FFA}_0$ [mM]           & 0.453        & 0.453         & 0.452      & 0.446      & 0.451      & 0.449      & 0.518 \\ \hline
$\text{TGL}_0$    [mM]        & 0.7      & 0.7       & 0.7       & 0.715     & 0.7       & 0.7       & 0.64 \\ \hline
$\text{GLR}_0$    [mM]        & 0.093    & 0.093     & 0.093     & 0.03      & 0.055     & 0.093     & 0.5 \\ \hline
$\text{KET}_0$    [mM]        & 0.0055   & 0.0073    & 0.0073    & 0.0092    & 0.0073    & 0.0066    & 0.0073 \\ \hline
$\text{PRO}_0$    [mM]        & 0        & 0         & 0         & 0         & 0         & 11541     & 0      \\ \hline
$\text{TGL}_{AP, 0}$ [mM]     & 0        & 0         & 0         & 0         & 0         & 0         & 5978   \\ \hline
$\text{ins}_0 [\frac{mU}{L}]$ & 15.1765  & 15.1765   & 15.1765   & 21.5542   & 10.6235   & 12.9      & 12.9 \\ \hline
$\text{glu}_0 [\frac{ng}{L}]$ & 100      & 100       & 100       & 98.8835   & 100      &    100    & 100  \\\hline

\end{tabular}%
\end{table}
\normalsize

\end{document}